\documentclass[twocolumn]{aa}  
\usepackage{graphicx}
\usepackage{natbib}
\bibpunct{(}{)}{,}{a}{}{;}
\usepackage[lofdepth,lotdepth]{subfig}
\usepackage{amssymb}
\usepackage{amsmath}
\usepackage{txfonts}

\begin{document}
\title{Minimizing follow-up for space-based transit surveys using full lightcurve analysis.}
\author{S.V. Nefs$^1$, I.A.G. Snellen$^1$ \& E.J.W. de Mooij$^1$}
\offprints{nefs@strw.leidenuniv.nl}
\institute{$^1$ Leiden Observatory, Leiden University, P.O. Box 9513, 2300 RA Leiden, The Netherlands}
\date{Revised Draft: 14 May 2012} \titlerunning{}%
\abstract
   {One of the biggest challenges facing large transit surveys is the elimination of false-positives from 
    the vast number of transit candidates. A large amount of expensive follow-up time is spent on verifying the nature 
    of these systems.}
   {We investigate to what extent information from the lightcurves can identify blend scenarios and eliminate them as planet candidates, to significantly decrease the amount of follow-up observing time required to identify the true exoplanet systems.}
   {If a lightcurve has a sufficiently high signal-to-noise ratio, a distinction can be made between the lightcurve of a stellar binary blended with a third star and the lightcurve of a transiting exoplanet system. We first simulate lightcurves of stellar blends and transiting planet systems to determine what signal-to-noise level is required to make the distinction between blended and non-blended systems as function of transit depth and impact parameter. Subsequently we test our method on real data from the first IRa01 field observed by the CoRoT satellite, concentrating on the 51 candidates already identified by the CoRoT team.}
   {Our simulations show that blend scenarios can be constrained for transiting systems at low impact parameters. At high impact parameter, blended and non-blended systems are indistinguishable from each other because they both produce V-shaped transits. About 70\% of the planet candidates in the CoRoT IRa01 field are best fit with an impact parameter of $b>$0.85, while less than 15\% are expected in this range considering random orbital inclinations. By applying a cut at $b<0.85$, meaning that $\sim$15\% of the potential planet population would be missed, the candidate sample decreases from 41 to 11. The lightcurves of 6 of those are best fit with such low host star densities that the planet-to-star size ratii imply unrealistic planet radii of $R>2R_{Jup}$. Two of the five remaining systems, CoRoT1b and CoRoT4b, have been identified as planets by the CoRoT team, for which the lightcurves alone rule out blended light at 14\% (2$\sigma$) and 31\% (2$\sigma$). One system possesses a M-dwarf secondary, 
one a candidate Neptune.}
{We show that in the first CoRoT field, IRa01, 85\% of the planet candidates can be rejected from the lightcurves alone, if a cut in impact parameter of $b<0.85$ is applied, at the cost of a $<15\%$ loss in planet yield. We propose to use this method on the Kepler database to study the fraction of real planets and to potentially increase the efficiency of follow-up. }   

   \maketitle

\section{Introduction}
With the CoRoT and Kepler space observatories in full swing (Baglin et al. 2006, Borucki et al. 2003), which both deliver thousands of lightcurves with unprecedented photometric precision and cadence, we have moved into an exciting new era of exoplanet research. Now, the characterisation of small, possibly rocky planets has finally become a realistic prospective (e.g. Corot-7b, Leger et al. 2009; Kepler-10b, Batalha et al. 2011). One of the biggest challenges is to seperate real planets from the significant fraction of (astrophysical) false-positives that can mimic a genuine transit signal (e.g. Batalha et al. 2010). Ground-based 
transit surveys have revealed that stellar eclipsing binaries (EBs) blended with light from a third star are the main source of contamination (e.g. Udalski et al. 2002). Also, for Super-Earth planet candidates blends with a background transiting Jupiter-sized planet system can be important. In these systems the eclipse depth, shape and ellipsoidal light variations of an EB are diluted by the effects of chance alignment of a foreground or background star or associated companion inside a photometric aperture set by either the pixel scale or the point spread function. In addition, light from a third star in the photometric aperture can bias the fitted parameters of a planet transit system. High resolution, high signal-to-noise spectra are normally required to exclude binary scenarios by excluding their large radial velocity or bi-sector variations, a process that can be very time-consuming.

Stellar blends are common in space-based transit surveys as apertures are relatively large (e.g. 19$"$x21$"$ for CoRoT), and target fields are crowded since the number of target stars is maximized in this way. To weed out false-positives, the CoRoT team relies on an extensive ground-based follow-up campaign for on-off photometry to identify the transited star in the CoRoT aperture (Deeg et al. 2009) and high resolution imaging observations to identify possible stars that dilute the lightcurve of a planet candidate. Even so, many candidates remain unresolved and defy easy characterisation after such a campaign. Kepler uses its unique astrometric precision to minimise the number of blends, which can be identified by a position shift of the flux centroid during transit, but will still require enormous ground-based efforts on the remaining $\sim$1200 candidates (e.g. Borucki et al. 2011). Together with the new influx of planet candidates from current surveys, possible future missions (such as PLATO; e.g. Catala 
et al. 2011) and ground-based efforts to hunt for planets around low-mass stars, the telescope demand for full follow-up may grow enormously. Therefore, any new technique or strategy that can eliminate even a moderate fraction of all candidates from the discovery lightcurves, prior to follow-up, is extremely valuable.

In this paper we investigate to what extent information from the lightcurves themselves can identify blend scenarios and eliminate them as planet candidates and on the other hand rule out blend scenarios in the case of true planet systems. Our key motivation is that \textit{the lightcurves of blended systems can not be perfectly fit by pure transit models and neither can genuine transits be fit by blended light models.} In section 2 we introduce our lightcurve fitting procedure and in section 3 we apply it to simulated data of a transiting hot Jupiter and Super-Earth. While such a procedure provides a natural tool to distinguish blends from genuine planetary systems by lightcurve fitting, it breaks down for transits with high impact parameters. We therefore only consider transiting systems with impact parameter $b<0.85$, loosing potentially $\sim$15\% of the planet catch, but significantly decreasing (by an order of magnitude) the required amount of follow-up observations. In section 4 we apply our method to 
the candidates of the CoRoT IRa01 field, whose candidates are almost completely characterised through an extensive follow-up campaign, and discuss the results in section 5.

\section{Method}
\subsection{Transit fitting} 
Several methods have been presented in the literature to identify blended systems and to select the best planet candidates. Seager \& Mallen-Ornellas (2003) proposed a diagnostic that involves fitting a trapezoid to the transit lightcurve to obtain estimates for the transit parameters and subsequently identify the best candidates. In this paper we use a method very similar to that used by Snellen et al. (2009) to reject blend scenarios for the transiting hot Jupiter OGLE2-TR-L9. It involves least-square fitting of a lightcurve using the standard transit parameters (see below) plus an additional parameter representing the extra light from a third light source. If the fit is significantly better with extra light, the lightcurve is from a blended system. If this is not the case, an upper limit to the third light fraction can be set to a degree depending on the signal-to-noise of the data. This procedure is in essence similar to \textit{Blender}, which is used by the Kepler team (e.g. Torres et al. 2011). 
However, \textit{Blender} simulates physical systems involving so many parameters that it is impractical to run on a large number of candidates. Here we are not interested in the true nature of the second object (whether it is a background, foreground or physically related star), just in its possible influence on the transit lightcurve.

We assume at this point that lightcurves with obvious signs of the presence of a stellar binary, such as ellipsoidal light variations and/or secondary eclipses, have been excluded from the candidate list. Note that a useful upper limit to the amount of ellipsoidal light variation, and the likelihood of a genuine planetary secondary, can be obtained by taking a Fourier transform of the data with the transit signal removed. We therefore do not require EBOP (Popper and Etzel 1981) to model the complex binary effects in the lightcurve, but rather utilize an IDL routine that incorporates the analytical transit model of Mandel and Agol (2002;M\&A). Our system simply consists of a secondary transiting a primary with possible additional light from a tertiary.

\subsection{Transit parameters}
We treat the transit mid-time $T_0$ and the orbital period $P$ as fixed parameters, resulting from the candidate selection process. For extra simplicity we keep the limb darkening parameters fixed at the tabulated solar values for CoRoT white light, assuming quadratic parameters (a,b)=(0.44,0.23) from Sing et al. (2010). Although this gives a small bias ($<$0.06 in impact parameter) for primary stars of different stellar type, the method is not meant for precise planet characterization and does not influence the characterization of potential blended and non-blended systems. Our transit model has three free parameters; the ratio of secondary over primary radii $(R_2/R_1)$, the impact parameter of the transit $b$, which is the smallest projected distance of the centre of the secondary to that of the primary in units of $R_1$, and the density of the primary star $\rho_1$. This density can be converted to the scaled orbital radius $(a/R_1)$, assuming that M$_1$$>>$M$_2$, through
\begin{equation}
\left(\frac{a}{R_1}\right)^3=\frac{G}{3\pi}\frac{\rho_1}{P^2}
\end{equation}
The relative projected distances $z$ between secondary and primary are computed from the input orbital phases $\phi$,
\begin{equation}
z(\phi)=\sqrt{\left(\frac{a}{R_1}\right)^2sin(\phi)^2+b^2cos(\phi)^2},
\end{equation}
Together with $(R_2/R_1)$, these are used as input to a custom-made IDL program, incorporating the routine from M\&A, that computes the theoretical models. We introduce light to this transit system by adding the blended light fraction $k$,
\begin{equation}
F_{total}(\phi,b,R_1/R_2,\rho_*,k)=F_{eclipse}\cdot(1-k)+k, 
\end{equation}
where $F_{eclipse}$ is the original transit lightcurve. We then devise the following chi-square statistic to compare the lightcurve to the data $F_{obs,i}$ with uncertainty $\sigma_{obs,i}$,
\begin{equation}
\chi^2=\sum_i\frac{(F_{obs,i}-F_{total,i})^2}{\sigma_{obs,i}^2}
\end{equation}
Note that we assume circular orbits. This has no influence on the characterization of blended and non-blended systems, but it does affect the derived host star density, and is therefore important for the estimate of the radius of the secondary object. This is further discussed in section 5.

\subsection{MCMC}
To obtain the best-matching system parameters, we use a Monte Carlo Markov Chain $\chi^2$ optimisation technique (MCMC, e.g. Tegmark et al. 1998) to map out the probability distribution for each lightcurve parameter. MCMC is found to be a more robust technique to obtain a global parameter solution in multi-parameter space than (downhill) grid-based methods, due to the resolution inefficiency of the latter (e.g. Serra et al. 2011). In the MCMC algorithm, the parameters $p_i$ are perturbed by an amount drawn from a normal distribution $\mathcal{N}$ according to: $p_{i+1}=p_i+f\cdot\mathcal{N}\cdot\sigma_p$, where $f$ is the jump function and $\sigma_p$ the standard deviation of the sampling distribution for each $p$. Subsequently $\chi^2$ is recalculated for these perturbed parameters and a Gaussian likelyhood $\mathcal{L}\varpropto exp(-\chi^2/2)$ is determined. These random jumps in parameter space are accepted or rejected according to the Metropolis-Hastings rule (Metropolis et al. 1953;Hastings 
1970)
. If the perturbed parameter set has a higher likelyhood $\mathcal{L'}$ than its progenitor, it will be accepted as a new chain point, otherwise it will be accepted with a probability of $\mathcal{L'}/\mathcal{L}$. We run the algorithm many times to build up a 'chain' of parameter values and tweak $\sigma_p$ and $f$ such that $\sim$40\% of the jumps are accepted. After creating multiple chains from different starting conditions, we check proper model convergence and mixing of the individual chains using the Gelman \& Rubin $R$ statistic (Gelman \& Rubin 1992). To save time, first $k$ is set to zero at the minimum $\chi^2$ determined with MCMC analysis. Subsequently $k$ is increased in small steps (but always kept fixed during the MCMC) with the previously found parameters as starting values. In this way the parameter values (adopting the median of the distribution) and the uncertainties in the parameters are determined as function of $k$ in an efficient way.
\section{Tests on synthetic lightcurves}
In this section, we test our method on synthetic lightcurves to determine the required precision to detect or exclude third light in a particular transit system.  We perform these simulations for two candidate systems: (i) a hot Jupiter orbiting a solar type star and (ii) a Super-Earth around a similar host. 
\subsection{Transiting hot Jupiter}
\begin{figure}[h!]
\centering
\includegraphics[width=0.5\textwidth]{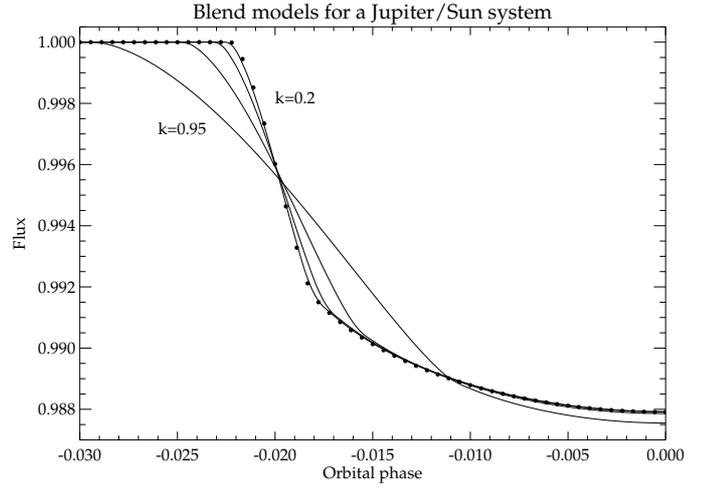}
\caption{\small{\textit{Simulated lightcurve for a transiting exoplanet system consisting of a hot Jupiter in a 2.5 day orbit around
a solar type star with impact parameter $b$=0.2 (black dots). The solid curves show diluted binary models with best-fit parameters determined by MCMC, for blended light fraction k=[0.2, 0.5, 0.8, 0.95].}}}
\label{jupiter}
\end{figure}

\begin{figure}[h!]
\centering
\includegraphics[width=0.5\textwidth]{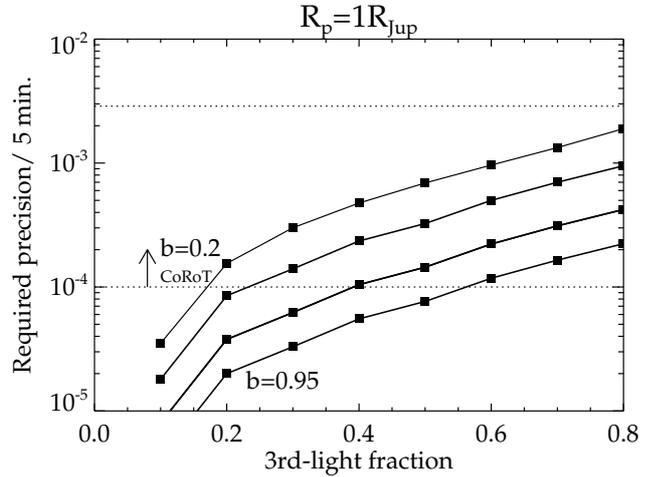}
\caption{\small{\textit{The photometric precision per 5 minutes required to exclude a blended light fraction k at 3$\sigma$ for a hot Jupiter around a solar type host star ($R_2/R_1=0.1$ and $\rho_*=\rho_{\odot}$), as a function of the system parameters $b$ and $k$. The four solid curves are for impact parameters b=0.2, 0.5, 0.8, and 0.95. The upper and lower horizontal dotted lines indicate the range in precision for objects in the IRa01 field, determined by Aigrain et al. (2009).}}}
\label{precision}
\end{figure}

We simulated a set of transit lightcurves for a hot Jupiter with $R_2=1\rm{R}_{\rm{Jup}}$ and P=2.5 days, orbiting a star with a solar density, for a range of impact parameters. The lightcurve for an impact parameter of $b=0.2$ is shown in Figure \ref{jupiter}. As explained in the previous section, our method finds the best fit for a range in blended light fraction $k$. Of course, in this simulation a perfect fit is obtained for $k$=0. As can be seen in Figure \ref{jupiter}, an increasingly worse fit is obtained for increasing $k$, most obviously seen by comparing the $k$=0.95 model to the synthetic data. This latter model fit assumes that 95\% of the light is from a third object, meaning that the unblended transit is actually a factor 20 deeper, hence 20\% deep instead of 1\%. It implies that $R_2/R_1\sim0.45$, resulting in a much longer duration transit unless it is grazing. This results in the best-fitting $k$=0.95 model being much more V-shaped than the synthetic lightcurve of the planet. We can now 
convert the differences between the synthetic lightcurves and model fits to $\chi^2$ values for each combination of $b$ and $k$ by assigning uncertainties to the synthetic data. In this way we can determine what photometric precision is required to exclude a certain blended light fraction in the lightcurves as a function of $b$. Figure \ref{precision} shows the precision per 5 minutes of data required to exclude a blended light fraction $k$ at a 3$\sigma$ level in a system with an impact parameter of b=0.2, 0.5, 0.8, and 0.95. The required precision becomes more stringent for lower values of $k$ and higher values of $b$. For $b$=0.2, 80\% blended light ($k$=0.8) can be excluded in a lightcurve with a precision of only $2\times10^{-3}$ per 5 minutes, while for $b$=0.8, 20\% of blended light can only be rejected if the lightcurve has a precision of $4\times10^{-5}$ per 5 minutes.
\subsection{Transiting super-Earth}
\begin{figure}[h!]
\centering
\includegraphics[width=0.5\textwidth]{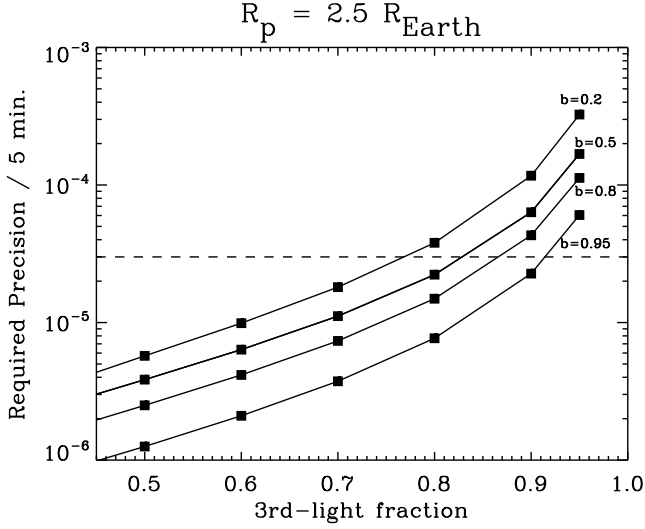}
\caption{\small{\textit{As for Fig. \ref{precision}, but for a 2.5$R_{Earth}$ SuperEarth planet around a solar type host in a 2.5 day orbital. We can exclude 80\% blended light at the 3$\sigma$ level at a moderate impact parameter of $b$=0.5. The horizontal dashed line refers to the precision reached in the discovery lightcurve of CoRoT 7b, the first rocky SuperEarth planet (Leger et al. 2009).}}}
\label{supearth}
\end{figure}

We performed also tests on a Super-Earth with $R_2=2.5R_{\oplus}$ orbiting a sun-like star, following the same procedure as described above. Since the transit itself is a factor $\sim16$ more shallow than for a Jupiter-size planet, the level of precision required to reject blend scenarios is also significantly higher, as can be seen in Figure \ref{supearth}. Note however that even for a blended light fraction of $k$=0.95, the radius of transiting object $R_2$ is still in the Jupiter-size regime. Hence only if the blended light fraction is very high, $k>0.95$, can an eclipsing binary mimic a Super-Earth transit.
\section{Tests on candidates in the CoRoT IRa01 field}
\subsection{The data set}
In this section we test our method on real data, using the lightcurves of the candidates selected by the CoRoT team from CoRoT field IRa01 (Carpano et al. 2009). In this first field targeted by CoRoT, 3898 bright stars were observed in chromatic mode (with a blue, green and red channel) and another 5974 in a single monochromatic "white" band in a 66 day staring run towards the Galactic anti-center. From the 50 initial candidates, a subsample of 29 promising targets received extensive follow-up as discussed in Moutou et al. (2009). Two of these have so far been identified as genuine planets: CoRoT-1b, a low density $R_p=1.49R_{Jup}$ transiting hot Jupiter around a G0V host (Barge et al. 2008) and CoRoT-4b, a $R_p=1.19R_{Jup}$ hot Jupiter around a F8V host (Aigrain et al. 2008). Seventeen additional systems were solved using the photometric and spectroscopic follow-up observations (Moutou et al. 2009). We choose to test our method on the 45 bright candidates with more than one transit observed, using the 
publicly available N2-level data.
\subsection{Pre-cleaning of the lightcurves}
We first combine the multicolor lightcurves into one single 'white lightcurve' for each candidate under the assumption that the CoRoT 
analysis teams did not detect any significant variation of eclipse depth with wavelength, which would already have been a clear sign of blending effects. We first clip each lightcurve by removing outliers at the 5$\sigma$ level. These outliers are mostly associated with the epochs at which the satellite passes the South Atlantic Anommaly (SAA)  or moves in/out of the Earth's shadow. We then iteratively refine the mid-times $T_0$ and the orbital period $P$ using the Kwee-van Woerden method (Kwee \& van Woerden 1956) and cross-correlation with a theoretical transit model (e.g. Rauer et al. 2009). Individual transit events that show temporary jumps in flux, caused by the impact of energetic particles (mainly protons) onto the CCD ("hot pixels"), are excluded from our analysis. For 16 out of the initial 50 CoRoT IRa01 candidates (32\%) we had to remove one or more transits from the lightcurve that were affected by such particle hits. Each individual lightcurve was then phasefolded around every 
transit. To normalise the data, we fit either a first order polynomial in a small range in phase ($\pm$0.1 from mid-transit) around each transit or a higher order polynomial (order $n$=13) in a larger phase range (typically $\pm$0.4 in phase), depending on which approach delivers the lowest rms in and out of eclipse and the least red noise (Pont et al. 2006). 
\begin{figure}[h!]
\centering
\includegraphics[width=0.5\textwidth]{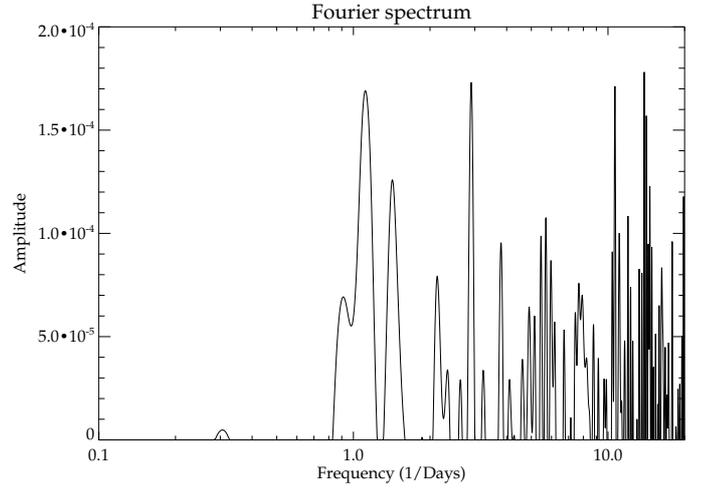}
\caption{\small{\textit{Fourier diagram of an example noise spectrum prior to lightcurve cleaning. Amplitude of the best-fitting sine curve on the vertical axis is plotted against frequency. Peaks around frequencies of 1.0 and $\sim$14 are due to remaining systematics related to the satellite orbit and Earth's rotational period.}}}
\label{Fourier}
\end{figure}
Figure \ref{Fourier} shows a typical example of the dominant frequencies still remaining after the polynomial fit. For most objects, distinct peaks exist around periods of 103 minutes and at 24 hours. We identify these peaks with remaining systematics, related to the satellite's orbit and Earth's rotational period, caused by ingress and egress of the spacecraft from Earth's shadow, variations in gravity and magnetic field and changes in the levels of thermal and reflected light from the Earth (e.g. Aigrain et al. 2009). By folding the out-of-eclipse data onto the dominant frequencies of the Fourier diagram, we then fit a sinusoidal function to the remaining systematics, followed by median averaging over all transits. We subsequently binned the lightcurves and assign errors, according to the standard deviation divided by the square root of the number of points in each bin.
\begin{figure}[]
\centering
\includegraphics[width=0.5\textwidth]{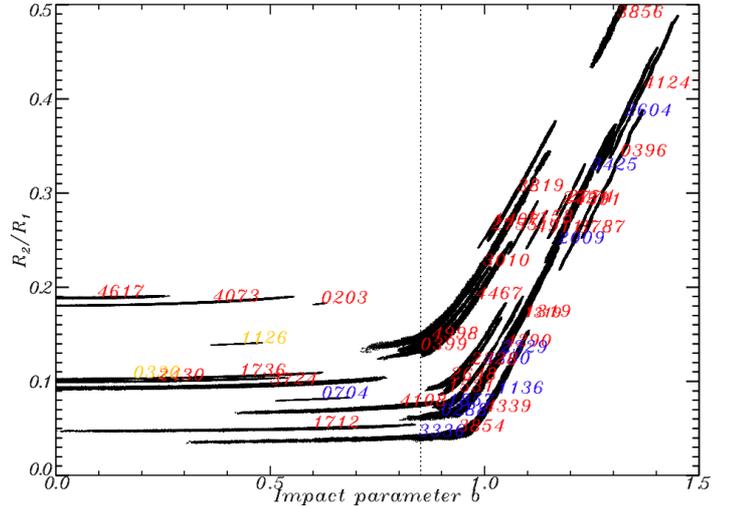}
\caption{\small{\textit{The MCMC solution for all IRa01 candidates in ($b,(R_p/R_*)$) space. Note the strong parameter degeneracy at high $b$. Yellow CoRoT WinIDs are the two confirmed planets CoRoT-1b and CoRoT-4b, blue objects are confirmed blends from the follow-up work presented in Moutou et al. (2009), and red sources are either unsettled cases or confirmed genuine binaries with non-planetary secondary masses from the radial velocity variations.}}}
\label{solutions}
\end{figure}
\subsection{Fitting the lightcurves}
\begin{figure}[h!]
\centering
\includegraphics[width=0.5\textwidth]{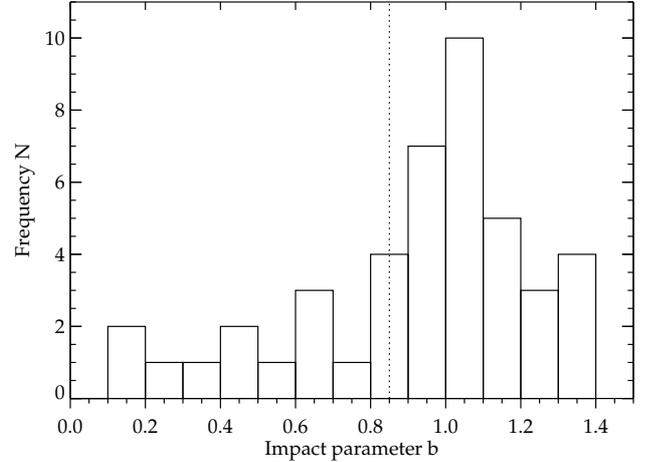}
\caption{\small{\textit{Distribution of fitted impact parameter of the CoRoT IRa01 candidates. The distribution is strongly peaked around b=1.0, indicating a significant population of (blended) EB contaminants. For a genuine planet distribution we would expect a flat histogram that falls off at high impact parameter. The dotted vertical line indicates the b=0.85 cutoff we have proposed in this paper.}}}
\label{histo}
\end{figure}

\begin{figure}[h]
\centering
\subfloat[][]{
\includegraphics[width=0.5\textwidth]{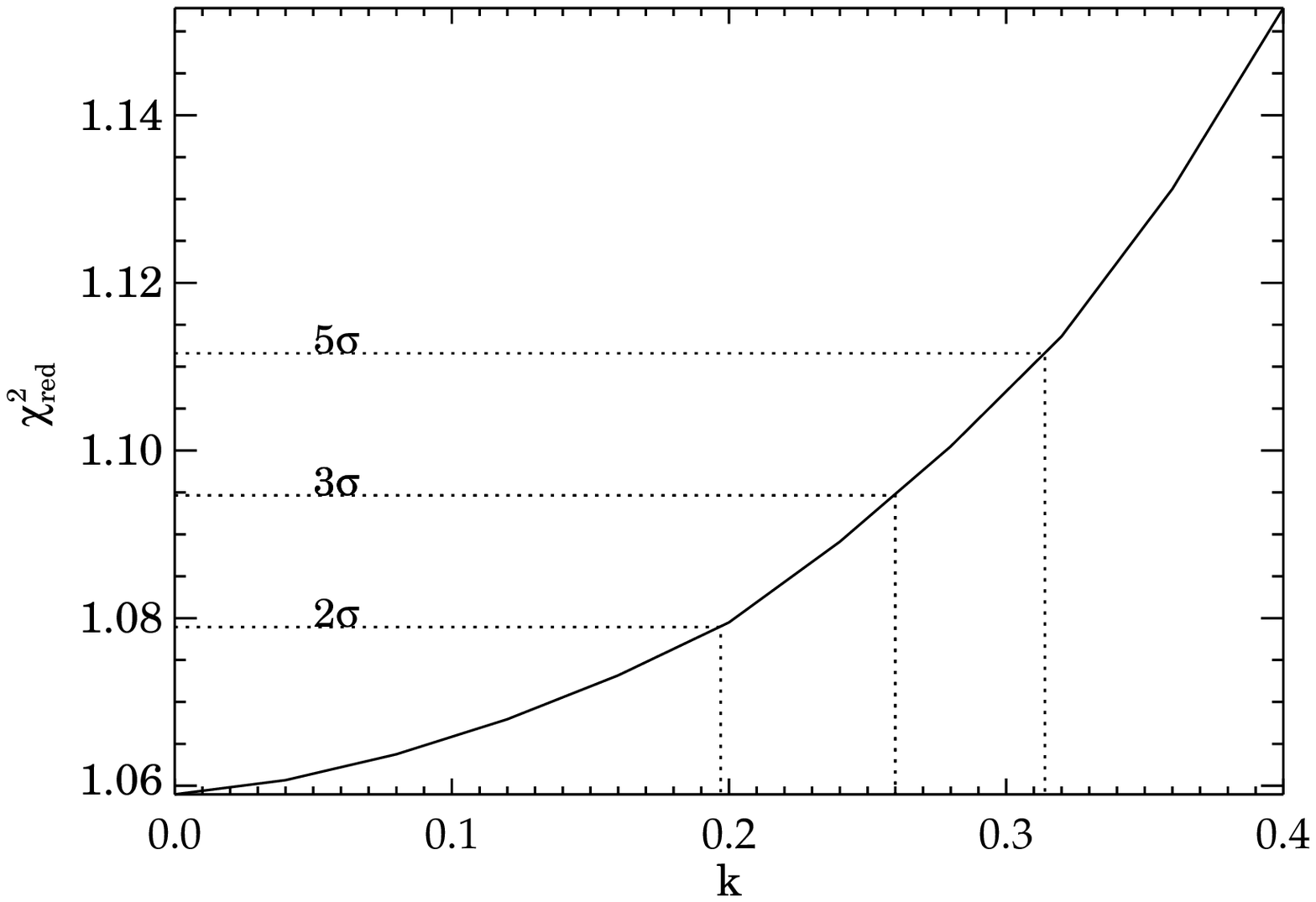}}
\qquad
\subfloat[][]{
\includegraphics[width=0.5\textwidth]{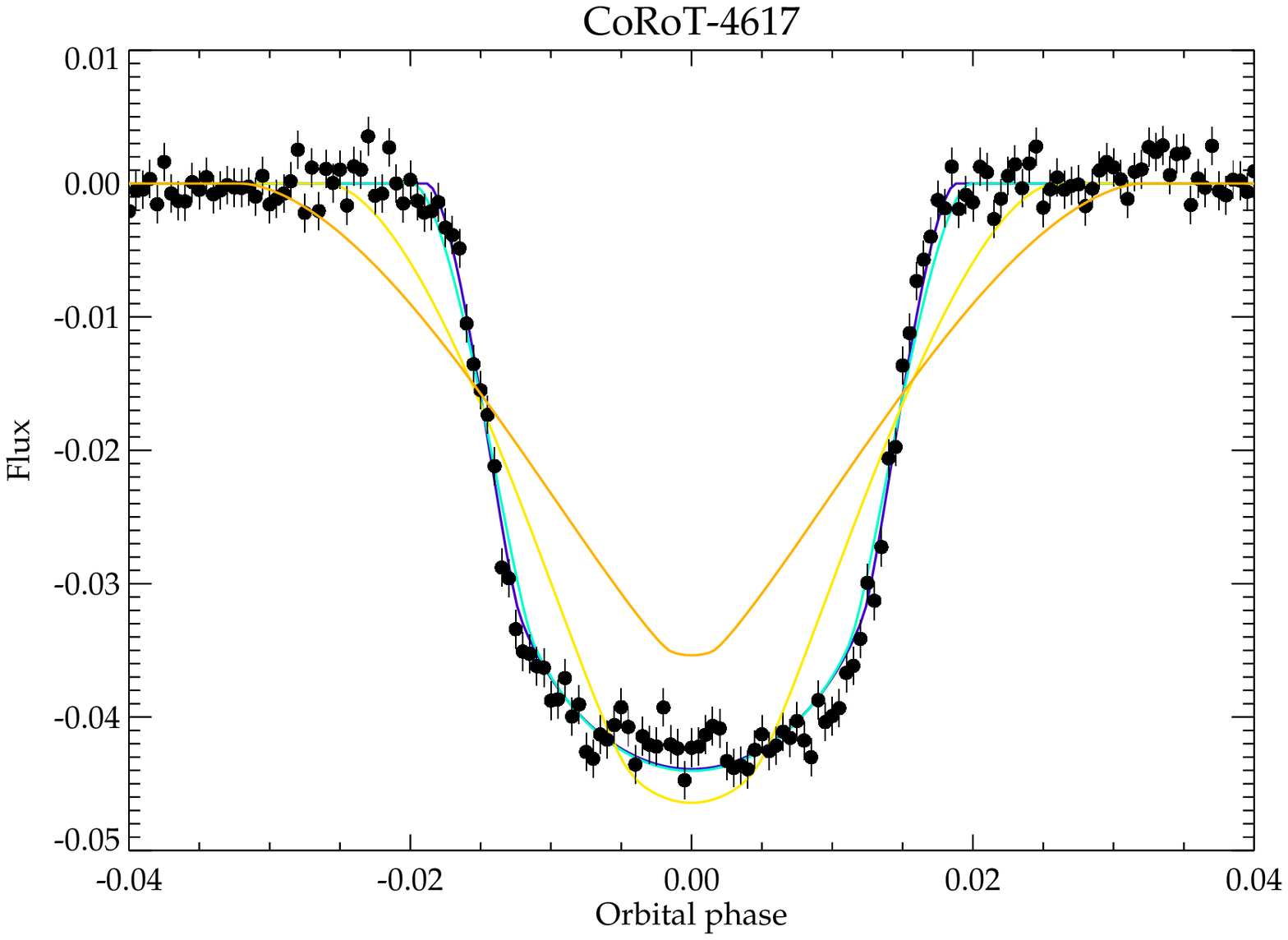}}
\caption{\small{\textit{Panel a): The reduced $\chi^2$ as function of blended light fraction $k$. The horizontal dashed line shows the 2,3 and 5$\sigma$ rejection criteria. Our lightcurve model directly indicates an early type main sequence stellar host, with a 2$\sigma$ upper limit for blended light of {\bf $k\sim20\%$}. The low stellar density implies a large secondary radius, rejecting the planet hypothesis. Panel b): best fitting EB models with blended light fraction k=[0.2,0.5,0.90,0.95], clearly showing that solutions with low k are favoured. Note that an orbit with an eccentricity of e=0.5, orientated in the right way, could increase the estimated stellar density to that of the Sun, and decrease R$_2$ to 2 R$_{Jup}$.  This ambiguity can be easily removed by taking a single spectrum of the star, resolving its spectral type.}}}
\label{example}
\end{figure}

Each lightcurve is first fitted with the method explained in section 2, assuming k=0, yielding the starting parameter sets ($R_2/R_1,b,\rho_*$) for our blend analysis. In Figure \ref{solutions} we show the resulting MCMC distribution of impact parameter $b$ versus $R_2/R_1$ for all the 45 candidates. CoRoT WinIDs (a shortcut of the CoRoT run identification number, e.g. IRa01-E1-2046) for each candidate are indicated, with yellow for the two confirmed planets CoRoT1b and CoRoT4b, in blue those candidates that have been confirmed to be blended systems by Moutou et al. (2009), and in red unsettled cases (either suspected early type stars with only few or very broad spectral lines for further radial velocity follow-up observations with HARPS or confirmed genuine EBs with non-planetary secondary masses). As can be seen, a large fraction of the candidates are, assuming no blended light, best fitted with a very high (often larger than unity) impact parameter. This is even more clear in the distribution 
of fitted impact parameters as shown in figure \ref{histo}. For 32 out of the 45 ($\sim70\%$) candidates $b>0.85$, while from geometric arguments it is expected that $\sim15\%$ of planets would be found at such a high impact parameter. Assuming that all eleven candidates at $b<0.85$ are non-blended systems only $\sim$1.6 objects are expected at $b>0.85$. Since our tests in section 3 have shown that it is very difficult to distinguish blends from non-blended systems at high impact parameters due to their V-shaped lightcurves, we apply a cut in the candidate list at $b<0.85$, knowing that we will potentially remove only a small fraction of the planet yield, in the case of the CoRoTa01 field $<0.3$ planets. From this it can be seen that it is highly likely that all candidates with $b>0.85$ are blended and/or grazing eclipsing binaries. For the eleven remaining candidates we used the transit parameters from the k=0 model to refit the lightcurve with an increasing value of $k$, as outlined in section 3. In this 
way we redetermine the best fit solution and $\chi^2$ as a function of $k$. As an example we show the best fit transit models for a range of $k$ and the $\chi^2$ as function of $k$ for candidate \textit{E1-4617} in Figure \ref{example}. As can be seen, the lightcurve can only be well fitted by models with a low $k$. E.g. the $\chi^2$ of the best fitting $k$=0.5 model is $\sim40\%$ higher than that for $k$=0. The 2 sigma upper limit for the fraction of blended light ($\Delta\chi^2$) is $k$=0.20. We performed this same analysis for all eleven remaining candidates for which the $\chi^2$ versus $k$ plots are shown in the Appendix, together with their best fit lightcurves. None of these candidates are better fitted by a high $k$ model than a low $k$ model, indicating that all blended systems have moved out of the remaining sample since they are all fitted with a high impact parameter. For six objects a significant fraction of blended light can be excluded from the lightcurve alone, including CoRoT-1b and CoRoT-4b.
 It would therefore not have been necessary to check whether the variability in these candidates came from the target star or not and the follow-up could have immediately concentrated on radial velocity measurements.

\begin{figure}[h!]
\centering
\includegraphics[width=0.5\textwidth]{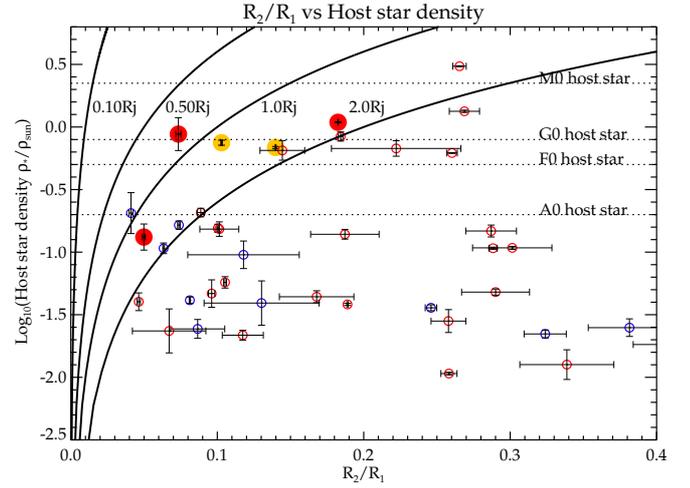}
\caption{\small{\textit{The $R_2/R_1$ size ratio versus the log of the stellar density for the CoRoT candidates in the IRa01 field, assuming k=0. The dotted lines mark the densities of A to M type main-sequence hosts. The five filled dots are the candidates that survive both our cuts in impact parameter and secondary size. The two confirmed transiting hot Jupiters CoRoT1b and CoRoT4b are shown as yellow filled dots. Open circles are the candidates we have excluded using our cuts. Blue circles indicate sources which have been identified as blended EBs by the CoRoT team follow-up, and red circles are either unsettled cases in the CoRoT follow-up or systems identified as genuine EBs through their radial velocities. The four solid curves indicate $R_2$=[0.10,0.50,1.0,2.0]$R_{Jup}$, assuming the main sequence mass-radius relation of Cox (2000) for the primary.}}}
\label{two}
\end{figure}

All parameters of the remaining candidates are shown in Table 1. An additional cut in the candidate list is made using a combination of the best fit mean stellar densities $\rho_*$ and $R_2/R_1$, as shown in Figure \ref{two}. Six of the candidates have host stars with densities corresponding to A-stars, resulting in unrealisticly large secondary radii of $>2.0R_{Jup}$. Note that there is currently no consensus on the upper limit of planet size, meaning that by setting a hard limit on planet radius we may exclude very large or bloated (hot) Jupiters. However, there are currently only 4 out of 219 transiting exoplanets reported with radii larger than 1.8$R_{Jup}$ (www.exoplanet.eu). Also, the probability that the secondary is a mid-type M-dwarf rather than a genuine planet increases when considering larger radii. This results in a remaining planet candidate sample of 5 objects instead of the original 45 using arguments based on the lightcurve alone. These five objects have been marked with filled symbols 
in Figure \ref{two}. Details on each system are discussed in Appendix A.
\begin{table}[!h]
  {
  \begin{tabular*}{0.5\textwidth}
      {@{\extracolsep{\fill}}ccccccc}
       \hline
   WinID+CoRoTID &P &$\left(\frac{R_2}{R_1}\right)$	&$b$	&$\textrm{Log}\left(\frac{\rho_*}{\rho_{\odot}}\right)$	&$\left(\frac{a}{R_*}\right)$ &2$\sigma$\\ 
       \hline \hline
   1126  0102890318&   1.51& 0.14& 0.43&-0.16& 4.93	&14\%\\
   0330  0102912369&   9.20& 0.10& 0.18&-0.13&16.96	&31\%\\
   0203  0102825481&   5.17& 0.18& 0.62& 0.04&13.09	&30\%\\
   1712  0102826302&   2.77& 0.05& 0.60&-0.88& 4.27	&93\%\\
   4108  0102779966&   7.37& 0.07& 0.80&-0.06&15.41	&95\%\\
   \hline\hline
   {{\tiny{\textit{$(R_2/R_1)$ versus $\rho_*$}}}}       &&&&&&\\
   4617  0102753331&  19.76& 0.19& 0.10&-1.42&10.47	&20\%\\
   2430  0102815260&   3.59& 0.10& 0.24&-0.81& 5.36	&44\%\\
   4073  0102863810&  15.00& 0.18& 0.36&-0.08&24.40	&67\%\\
   1736  0102855534&  21.72& 0.11& 0.43&-1.24&12.77	&62\%\\
   3724  0102759638&  12.33& 0.10& 0.50&-1.33& 8.17	&78\%\\
    \hline
\end{tabular*}
    }
  \label{re}
  \caption{\textit{The candidate sample that survives the impact parameter cut. The last six sources are excluded using a second cut because the fitted host star density indicates a secondary radius $R_2>2R_{Jup}$. The last column indicates the $2\sigma$ upper limit to the blended light fraction k.}} 
  \end{table}

\section{Discussion}
In this paper we investigated to what extent we can use the high signal-to-noise lightcurves of space-based exoplanet transit surveys to identify blended light scenarios, and eliminate them as planet candidates. We concentrated on the 51 exoplanet candidates from the first CoRoT IRa01 field (Carpano et al. 2009). About 70\% of the 51 planet candidates in the CoRoT IRa01 field are best fit with an impact parameter of $b>0.85$, which at face value already indicates that the candidate distribution is strongly contaminated by blended and/or grazing systems. We find that by cutting a candidate sample such that those objects with high impact parameter are removed, at the cost of losing a small fraction of potential planets, a significant reduction in required follow-up observations can be achieved. Of all candidates, only 5 remain in the final sample of which two are genuine planet systems, one is a low mass transiting M dwarf and one is a candidate Neptune.

The V-shaped lightcurves of near-grazing planet systems are strongly degenerate with blended eclipsing binary systems and can therefore not be distinguished from each other. How many planets are potentially missed by invoking the cut in impact parameter? Of the known transiting exoplanets, $\sim$6\% has an impact parameter larger than 0.85 and $\sim$16\% an impact parameter of more than 0.75\footnote{$www.exoplanet.eu$}. The cumulative probability of a particular transit at a given impact parameter greater or equal to a cutoff value $b_X$ and transit depth $\Delta F$ is given by:
\begin{equation}
P_c(b>b_X)=\frac{1+\sqrt{\Delta F}-b_X}{1+\sqrt{\Delta F}}=\frac{1+R_2/R_1-b_X}{1+R_2/R_1}
\end{equation}
Note that this expression is different from the equation presented in Seager and Mallen-Ornellas (2003), because the maximum impact parameter in their formula is determined by the grazing condition $b_{max}=1-R_2/R_1$, yielding a minus sign in equation 5. For a 1$R_{Jup}$ planet around a solar type star $\sim$22\% would potentially be missed by setting the cut in impact parameter ($\sim$6\% according to Seager and Mallen-Ornellas). However, extremely grazing systems will be very shallow and of short duration and will therefore provide very limited physical information. For example, a grazing 1$R_{Jup}$ with impact parameter $b=$1.05, will show a transit with a duration of 30\% and only 20\% of the depth of a transit with $b=0$. Therefore, the actual planet loss fraction will be closer to the predictions of Seager and Mallen-Ornellas (2003), i.e. $<15\%$. 

In this paper we have made the assumption of circular orbits, but radial velocity surveys teach us that such an assumption is not valid for longer periods (e.g. Butler et al. 2006). In addition, Barnes (2007) shows that a planet with an eccentric orbit is more likely to transit by a factor of $(1-e^2)^{-1}$ than a planet in a circular orbit with the same semi-major axis. A significant population of transiting exoplanets with an eccentric orbit is therefore expected for long duration space-based surveys. Because the planet orbital velocity varies from $\sqrt{\frac{1+e}{1-e}}V_{circ}$ to $\sqrt{\frac{1-e}{1+e}}V_{circ}$ between periastron and apastron in an eccentric orbit, transit duration can vary as function of $e$ and $\omega$ (the angle of pericenter). This leads to a wrong fit of the host star density (e.g. Kipping 2010a, Tingley et al. 2011), therefore directly affecting our estimate of the secondary radius $R_2$. We therefore can not reliable make the planet-to-star ratio versus host star density cut 
in the eccentric orbit case for longer period planets ($P>3.0$days). Fortunately, the fitted impact parameter, $R_2/R_1$ and blended light fraction $k$ are not affected by an eccentric orbit. This means that we can still first apply a cut in impact parameter $b<0.85$ and remove likely blends. To subsequently determine the real host star density it is sufficient to take a single high-resolution spectrum to determine $\rho_1$ and estimate $R_2$. Using this spectroscopically determined density an upper limit to $ecos(\omega)$ can be set. One particular case in our sample is CoRoT-4617 with an orbital period of P=19.76 days. Assuming a circular orbit, the host star is estimated to have a density only $\sim$4\% of that of the Sun, in accordance with an early B-star. This would imply that the radius of the secondary object has R$_2$$\sim$8R$_{Jup}$. However, an orbit with an eccentricity of e=0.5, orientated in the right way, could increase the estimated stellar density to that of the Sun, and decrease R$_2$ to 2 
R$_{Jup}$.  This ambiguity can be easily removed by taking a single spectrum of the star, resolving its spectral type.

The method presented here is designed to remove false-positives in candidate lists through the identification of blended light. We do not assign a likelihood of planetary nature to the remaining candidates, meaning that we do not assess whether these are genuine planet systems, we just removed those systems which are not (except for a small fraction of collateral damage). However, it is anyway interesting to link blended light fractions to the population of random background eclipsing binaries. Assuming that ~1:300 of field stars are eclipsing binaries (Devor et al. 2008), and 1:1000 stars have a transiting hot Jupiter, we require an average of 0.3 background stars within the PSF, and within the magnitude range set by the limit of blended light, to have an equal likelihood for the two scenarios, and to end up with half of the remaining objects as eclipsing binaries. For a typical magnitude (V=14.0) for the candidate star, taking into account the size of the CoRoT PSF, this is reached at a $\Delta$mag = 
$\sim$1.5, corresponding to k=0.8. For 8 out of the 10 remaining targets this level of blended light is excluded at the $>$3$\sigma$ level. For a typical limit of k$<$0.6, the background eclipsing binary can at most be 0.5 magnitudes fainter than the target star, making this scenario a factor $\sim$5 less likely. Do note however that this does not take into account physical triple systems, for which radial velocity follow-up is required to exclude them. Recently, the Kepler team have announced the discovery of $\sim$1200 planet candidates (Burucki et al. 2011). We propose to use the method presented here on this candidate list, to identify clear blend systems to obtain a better estimate on the fraction of planet systems in this sample.

\section{Conclusions}
In this paper we have investigated to what extent information from lightcurves of a space-based exoplanet transit survey can identify blended light scenarios and eliminate them as planet candidates, to significantly decrease the required amount of follow-up time. If a lightcurve has sufficiently high signal-to-noise, a distinction can be made between a blended eclipsing binary and a transiting exoplanet. We first have simulated lightcurves of stellar blends and transiting planet systems to determine the required signal-to-noise as a function of impact parameter and transit depth. Our simulations show that blend scenarios can be distinguished from transiting systems at low impact parameter. At high impact parameter, blended and non-blended systems both produce V-shaped transits and are indistinguishable from each other. We have subsequently tested our method on real data from the first IRa01 field of CoRoT, concentrating on the 51 candidates already identified by the CoRoT team (Carpano et al. 2009). We show 
that 70\% of the planet candidates in the CoRoT IRa01 field are best fit with an impact parameter of $b>0.85$, whereas $\sim$15\% are expected assuming random orbital orientations. By applying a cut at $b<0.85$, meaning that $\sim$15\% of the potential planet population would be missed, the candidate sample decreases from 41 to 11. The lightcurves of 6 of those are best fit with such a low host star density that the planet-to-star size ratio implies an unrealistic planet radius of $R_2>2\rm{R}_{\rm{Jup}}$. From the remaining five, two systems, CoRoT-1b and CoRoT-4b, have been identified by the CoRoT team as planets, for which the lightcurves alone rule out blended light at a 14\%(2$\sigma$) and 31\%(2$\sigma$). One other candidate is also consistent with a non-blended system, but is a late M-dwarf, which will always require radial velocity follow-up for confirmation since M-dwarfs can have similar radii as Jupiter mass planets. One other system consists of a candidate Neptune around a M-dwarf according to 
Moutou et al. (2009). We have therefore shown that 85\% of the planet candidates can be rejected for the IRa01 field from the lightcurves alone. We propose to use this method on the Kepler database to study the fraction of real planets and to potentially increase the efficiency of follow-up. For long period candidates, possible non-zero eccentricity will affect the cut in planet-to-star ratio versus host star density, effectively increasing the sample size. However a single high-resolution spectrum would be sufficient to determine the real host star density and estimate the size of transiting objects.

\pagebreak \clearpage
\section*{Appendix}
In this Appendix we discuss in detail the sample of 10 remaining CoRoT candidates, that were selected using the cut in impact parameter and were presented in Section 4 and Table 1. In Figures 9-11, we show for each candidate the blended light fraction $k$ versus reduced $\chi^2$, and the best fitting blended light models for k=0.2, 0.5, 0.9, and 0.95. In Table 2, we show best-matching system parameters for the full CoRoT IRa01, assuming no blended light.

\subsection*{Comments on individual sources}
\noindent {\bf{SELECTED PLANET CANDIDATES FROM THE LIGHTCURVES ALONE}}\newline
\textit{E2-1126-0102890318}\newline
We find a 2$\sigma$ upper limit for blended light contribution of {\bf $k<0.14$}, therefore a blend scenario can be excluded for this source at high confidence using the lightcurve alone. In addition, by assuming that the host star is on the main sequence, its mean density points to a $\sim$1.5$\rm{R}_{\rm{Jup}}$ radius, well in the range of known hot Jupiters. Of course, this source is exoplanet CoRoT-1b (Barge et al. 2008).\newline
\textit{E1-0330-0102912369}\newline 
We find a 2$\sigma$ upper limit for blended light contribution of {\bf $k<0.31$} from its lightcurve, meaning that only a small contribution of blended light is tollerated. Assuming the host star is on the main sequence, its mean density points to a $\sim$1.2$\rm{R}_{\rm{Jup}}$ radius for the secondary. This object is identified as exoplanet CoRoT-4b (Aigrain et al. 2008). Eventhough the CoRoT-4b host star is of similar brightness as CoRoT-1b, the significantly longer orbital period, the residual variability (caused by a spotted rotating stellar photosphere) and the 1.8 times smaller transit depth are the causes of the lower confidence on blended light.\newline
\textit{E2-0203-0102825481}\newline 
The 2$\sigma$ upper limit for blended light is {\bf $k<0.3$} from its lightcurve. Radial velocity follow-up by the CoRoT team showed this to be an eclipsing binary of a low-mass M dwarf and a G-type primary (Morales et al., in prep). Assuming the host star is on the main sequence, its mean density points to a $\sim$1.7$\rm{R}_{\rm{Jup}}$ radius. Although not a planet, it is consistent with a non-blended system as found from our lightcurve fitting. Such systems always require RV follow-up since late M dwarfs and Jupiter-mass planets can have similar radii.
\newline 
\textit{E2-1712-0102826302}\newline
We find a 2$\sigma$ upper limit for blended light contribution of {\bf $k<0.93$}. We can therefore only exclude a high contribution of blended light for this shallow (2.4 mmag) transit. This means that at 2$\sigma$ confidence the true eclipse depth is less than 2.4\% in the presence of blended light. The fitted host star mean density points to an early type or evolved system. HARPS radial velocity follow-up has confirmed that the host star is an evolved fast rotator and Moutou et al. (2009) conclude that a triple system is the most probable scenario.
\newline
\textit{E1-4108-0102779966}\newline 
Because of the poor signal-to-noise of this transit and the relatively high impact parameter $b=$0.8, the 2$\sigma$ upper limit for blended light is $k<0.95$, therefore only a very high contribution of blended light can be excluded for this candidate. Assuming the host star is on the main sequence, its density is slightly lower compared to the solar value, indicating a stellar radius of $R_1\sim1.2R_{\odot}$. However, spectroscopic follow-up with HARPS suggested that the host is a low mass ($\sim0.8M_{\odot}$) star. No additional follow-up has thusfar been obtained by the CoRoT team.

\noindent {\bf{CANDIDATES REJECTED DUE TO THEIR LARGE SIZE}}\newline
\textit{E1-4617-0102753331}\newline
The 2$\sigma$ upper limit for blended light is $k<0.20$, therefore a blend scenario can be excluded at high confidence for this source. Assuming the host star is on the main sequence, its very low density points to an early B-type primary with a K dwarf secondary. The planet hypothesis is rejected and no additional follow-up is therefore required judging from the lightcurve alone. Note that an orbit with an eccentricity of e=0.5, orientated in the right way, could increase the estimated stellar density to that of the Sun, and decrease R$_2$ to 2 R$_{Jup}$.  This ambiguity can be easily removed by taking a single spectrum of the star, resolving its spectral type.
\newline
\textit{E2-2430-0102815260}\newline
We find a $2\sigma$ upper limit for blended light contribution of $k<0.44$. Again, only a small contribution of blended light is tollerated. Assuming the host star is on the main sequence, its mean density, consistent with an A type or evolved star, points to a radius $R_2>2.5\rm{R}_{\rm{Jup}}$. Radial velocity follow-up by the CoRoT team showed this to be a single lined eclipsing binary of a fast rotating host star and an early type M dwarf (Moutou et al. 2009).
\newline
\textit{E2-4073-0102863810}\newline
For this source, we find a $2\sigma$ upper limit for blended light of {\bf {$k<0.67$}}. This object shows $\sim$4\% deep eclipses around a host star that is $\sim$20\% less dense than the sun. This candidate was introduced in the original list of Carpano et al. (2009), but is not mentioned in the follow-up paper of Moutou et al. (2009). With an anticipated secondary radius of $\sim$2.1$R_{Jup}$ this object could still belong to the rare group of low mass stars or brown dwarfs. In the case of a stellar M5 secondary, the secondary eclipse would be detectable at $\sim$3.5 mmag in depth.
\newline
\textit{E2-1736-0102855534}\newline
The 2$\sigma$ upper limit for blended light is $k<0.62$. The low mean density of the host star, consistent with a very early main sequence or evolved star, points to a $>2.0\rm{R}_{\rm{Jup}}$ radius. Analysis of the lightcurve reveals a secondary eclipse at the 9$\sigma$ level, which indicates the secondary is in fact a low mass star. CoRoT radial velocity follow-up has confirmed that the host star is a fast rotating early type star and the system is a single lined eclipsing binary.
\newline
\textit{E2-3724-0102759638}\newline
For this source, we find a $2\sigma$ upper limit for blended light of $k<0.78$. Assuming the host star is on the main sequence, its very low density points to an A type primary, therefore $R_2>2.0R_{Jup}$. This object is listed both as a planet candidate and a binary by Carpano et al. (2009).

\onecolumn
\begin{figure}[h!]
\centering
\subfloat[][CoRoT-1b]{
\includegraphics[width=0.45\textwidth]{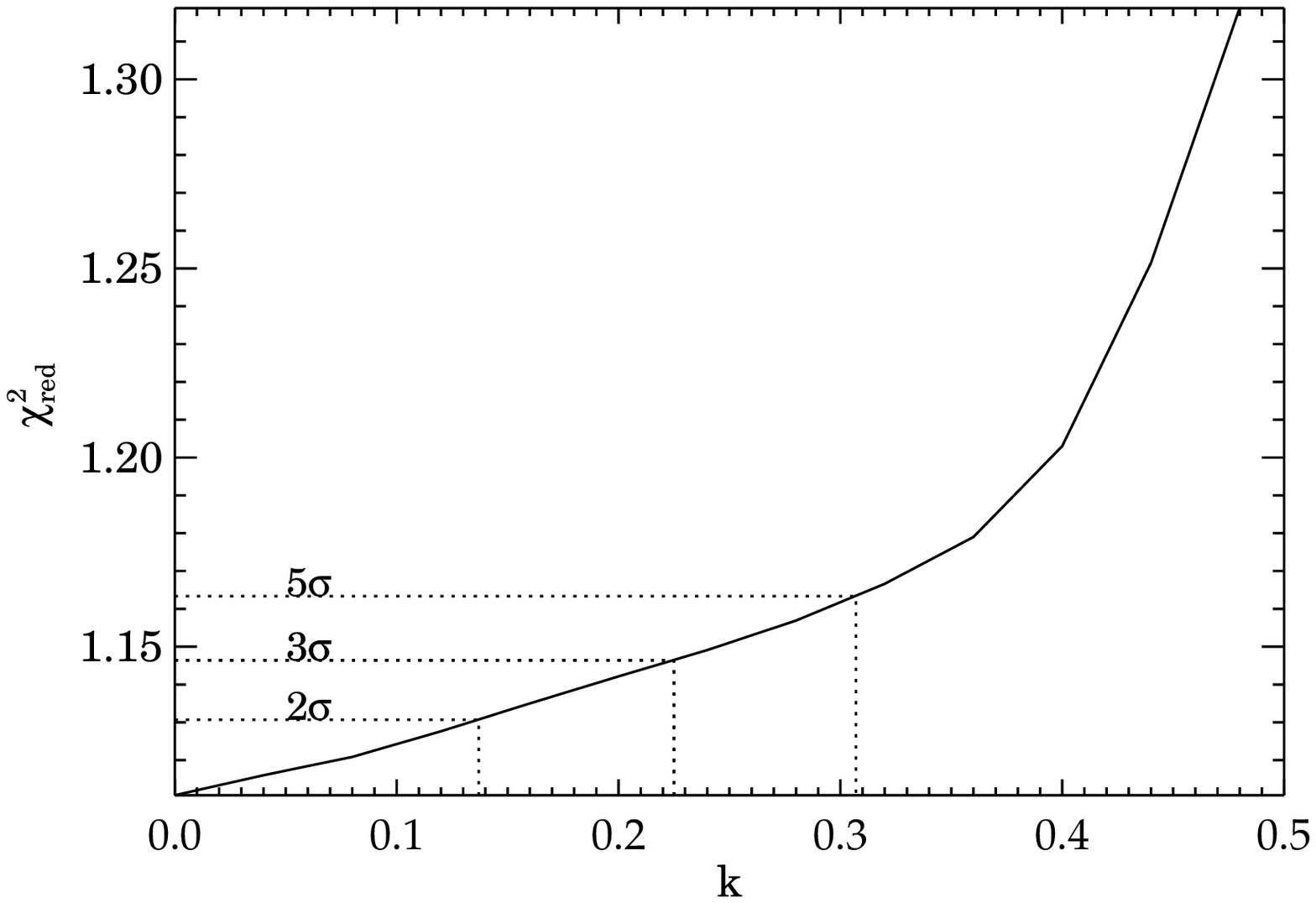}
\includegraphics[width=0.45\textwidth]{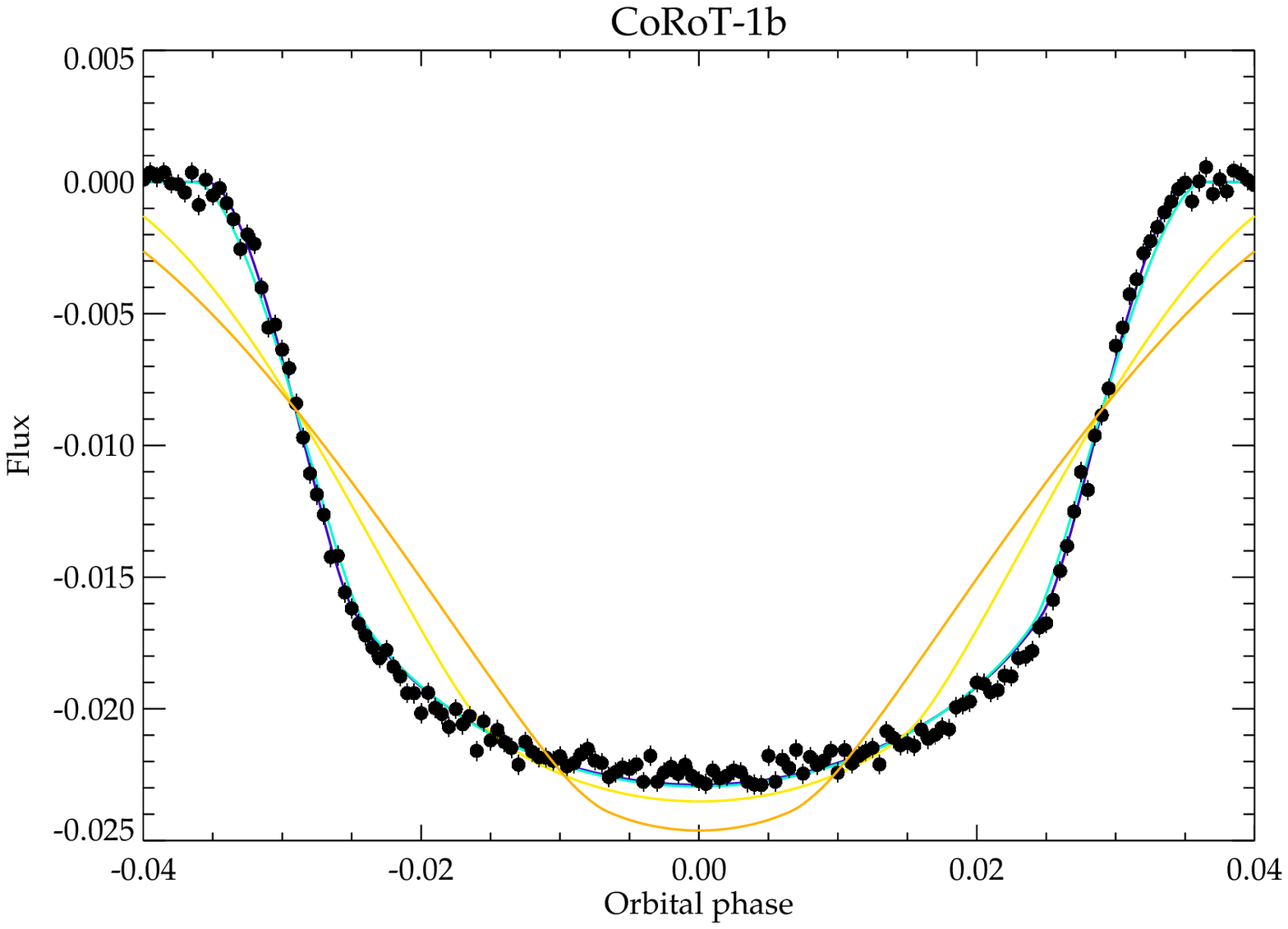}}
\qquad
\subfloat[][CoRoT-4b]{
\includegraphics[width=0.45\textwidth]{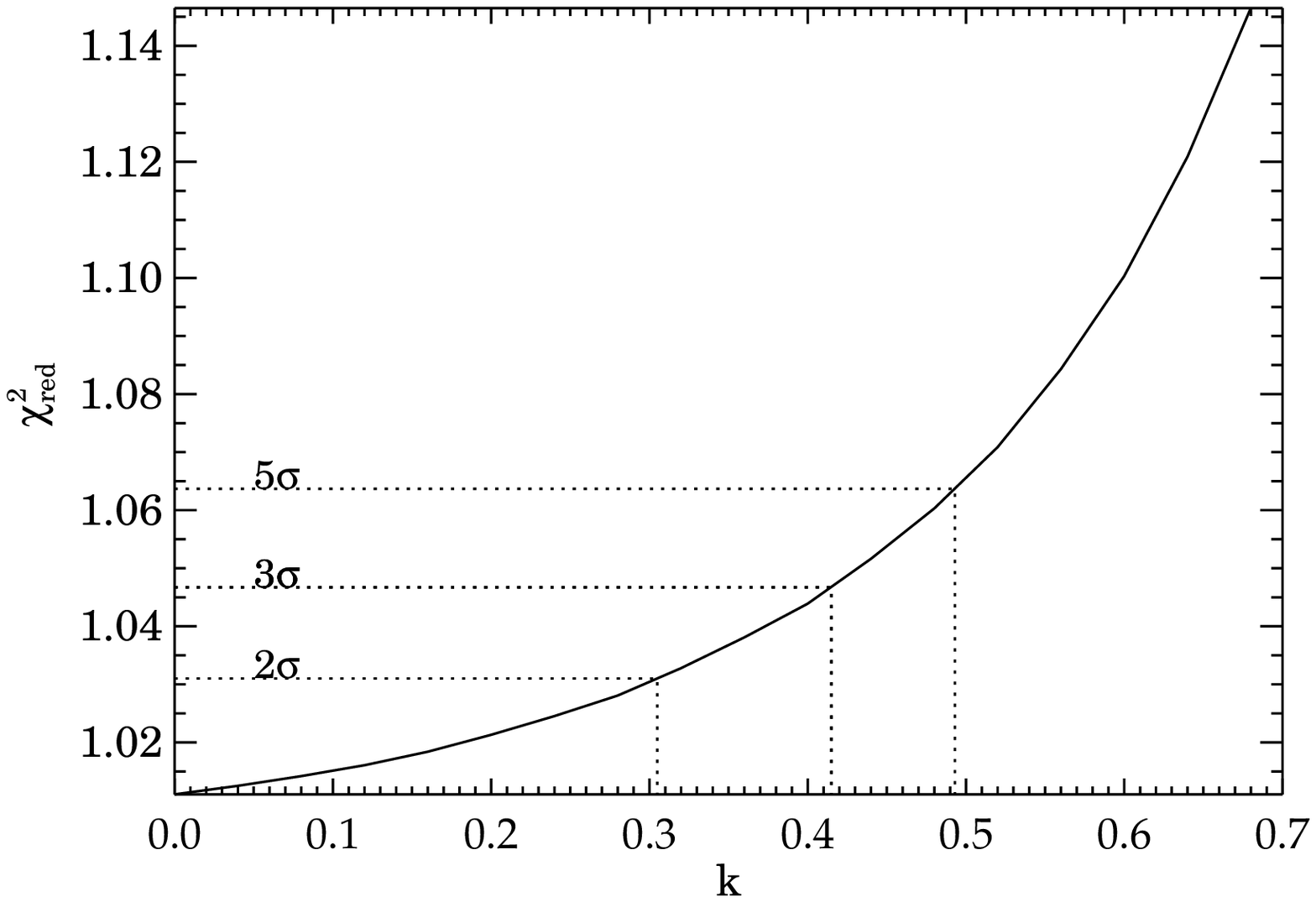}
\includegraphics[width=0.45\textwidth]{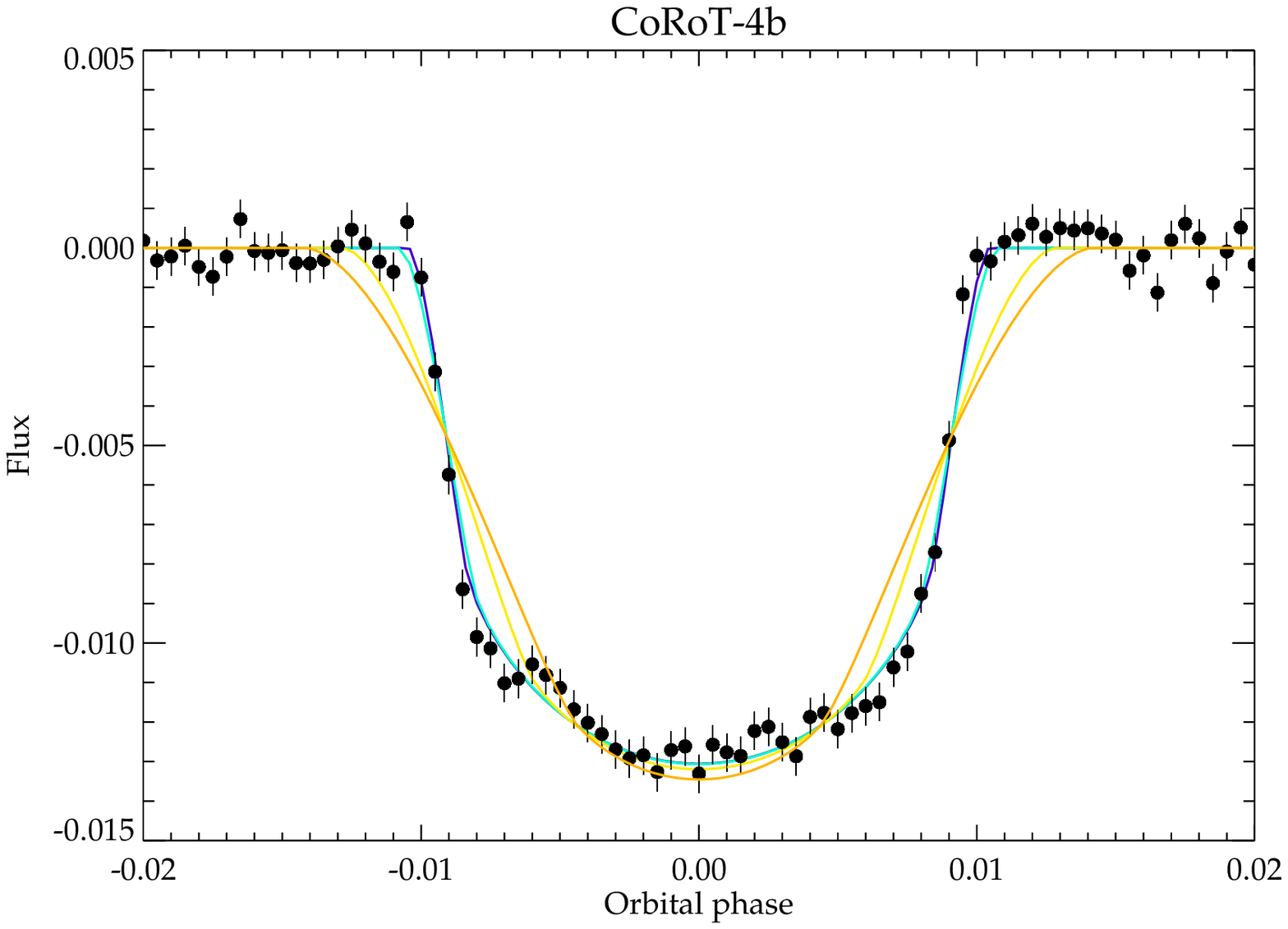}}
\qquad
\subfloat[][CoRoT-0203]{
\includegraphics[width=0.45\textwidth]{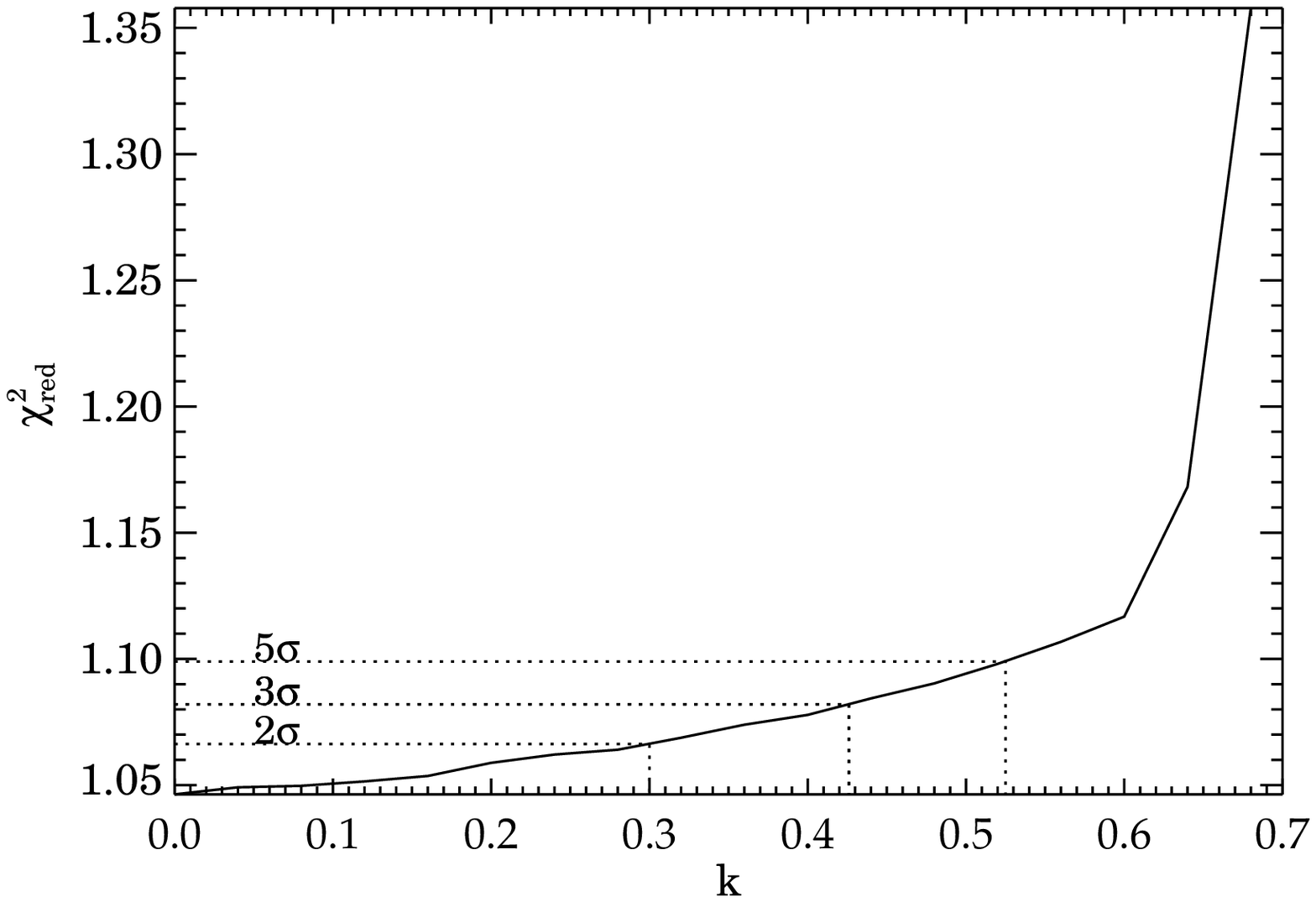}
\includegraphics[width=0.45\textwidth]{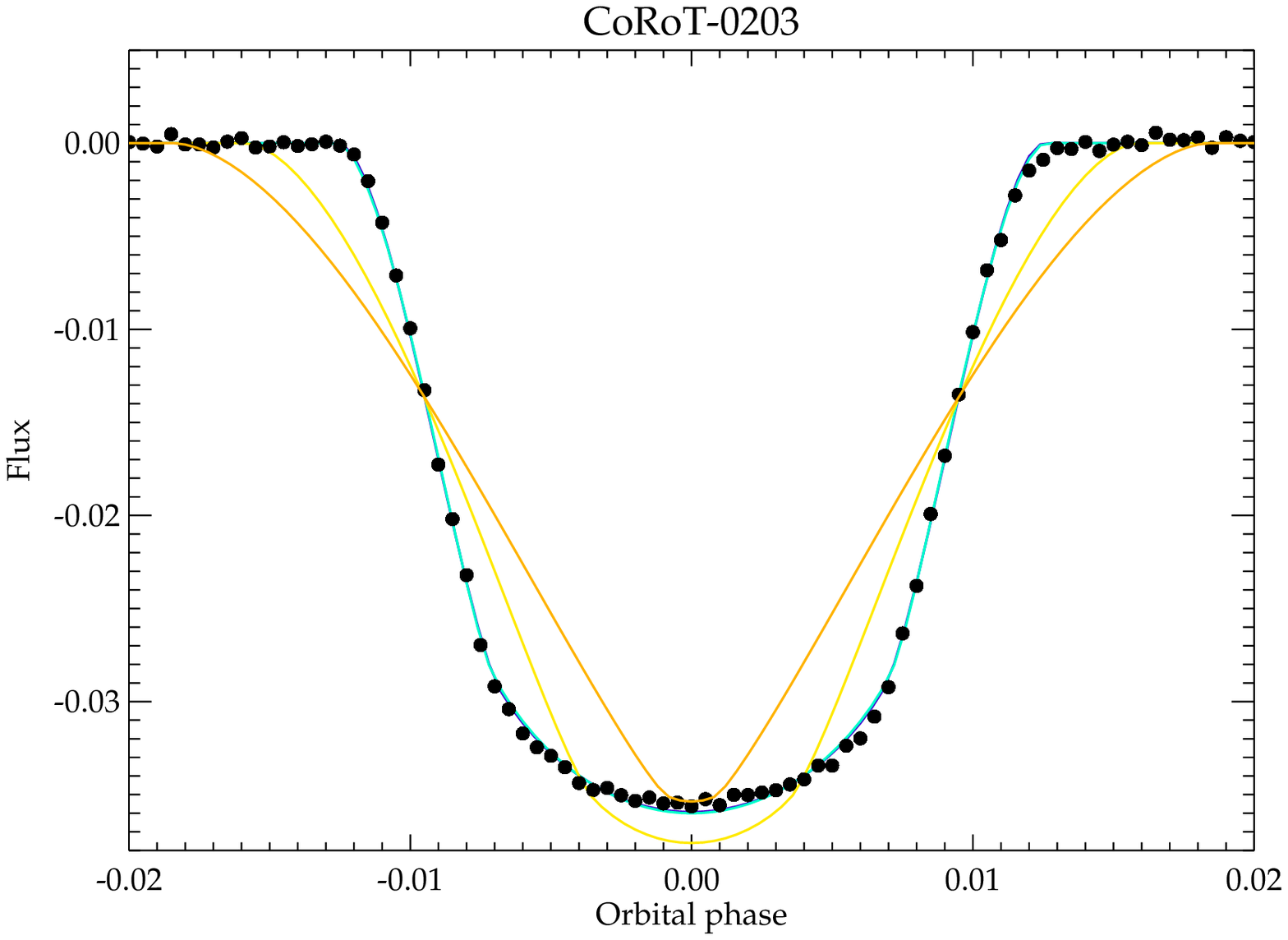}}
\qquad
\subfloat[][CoRoT-1712]{
\includegraphics[width=0.45\textwidth]{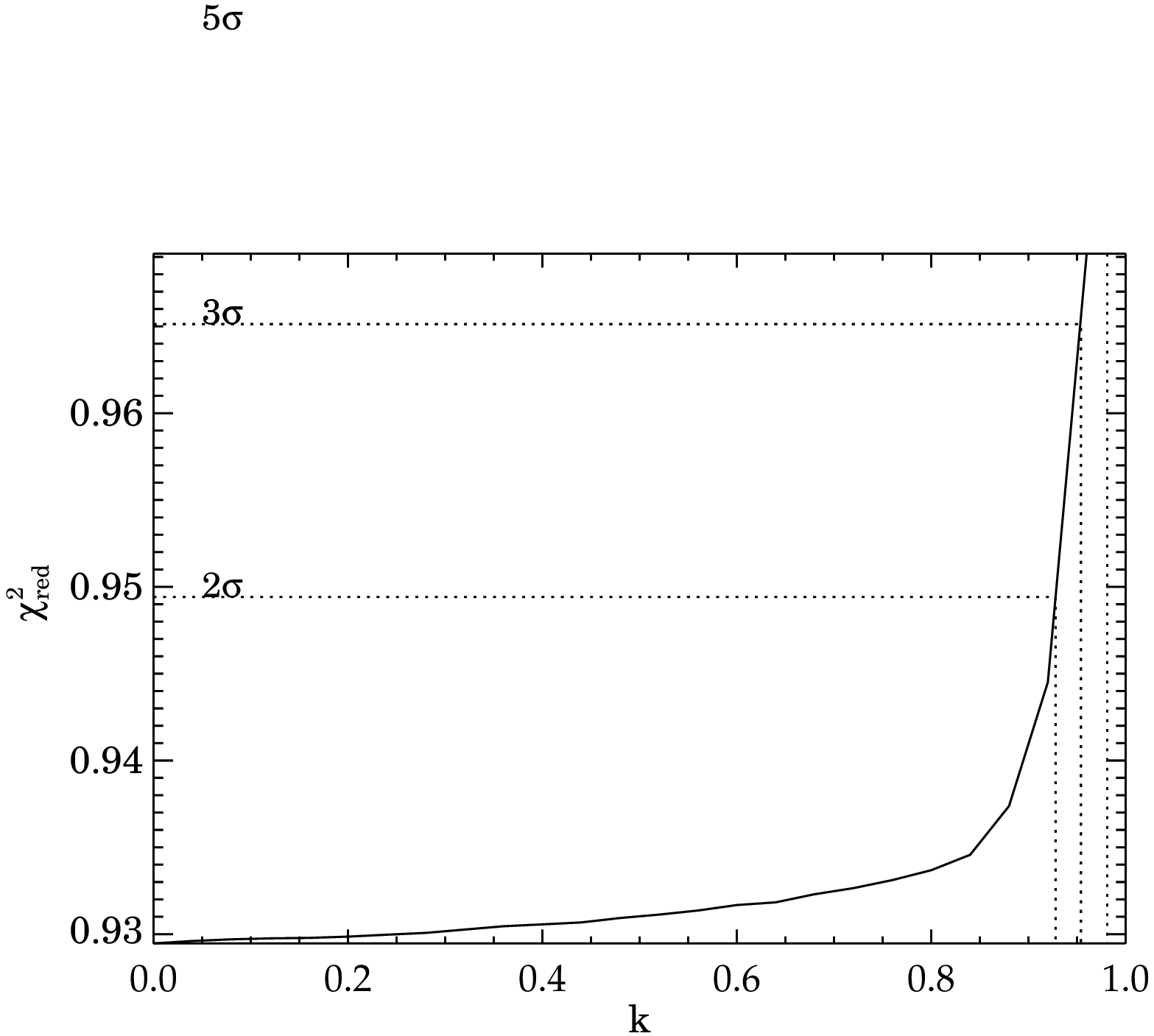}
\includegraphics[width=0.45\textwidth]{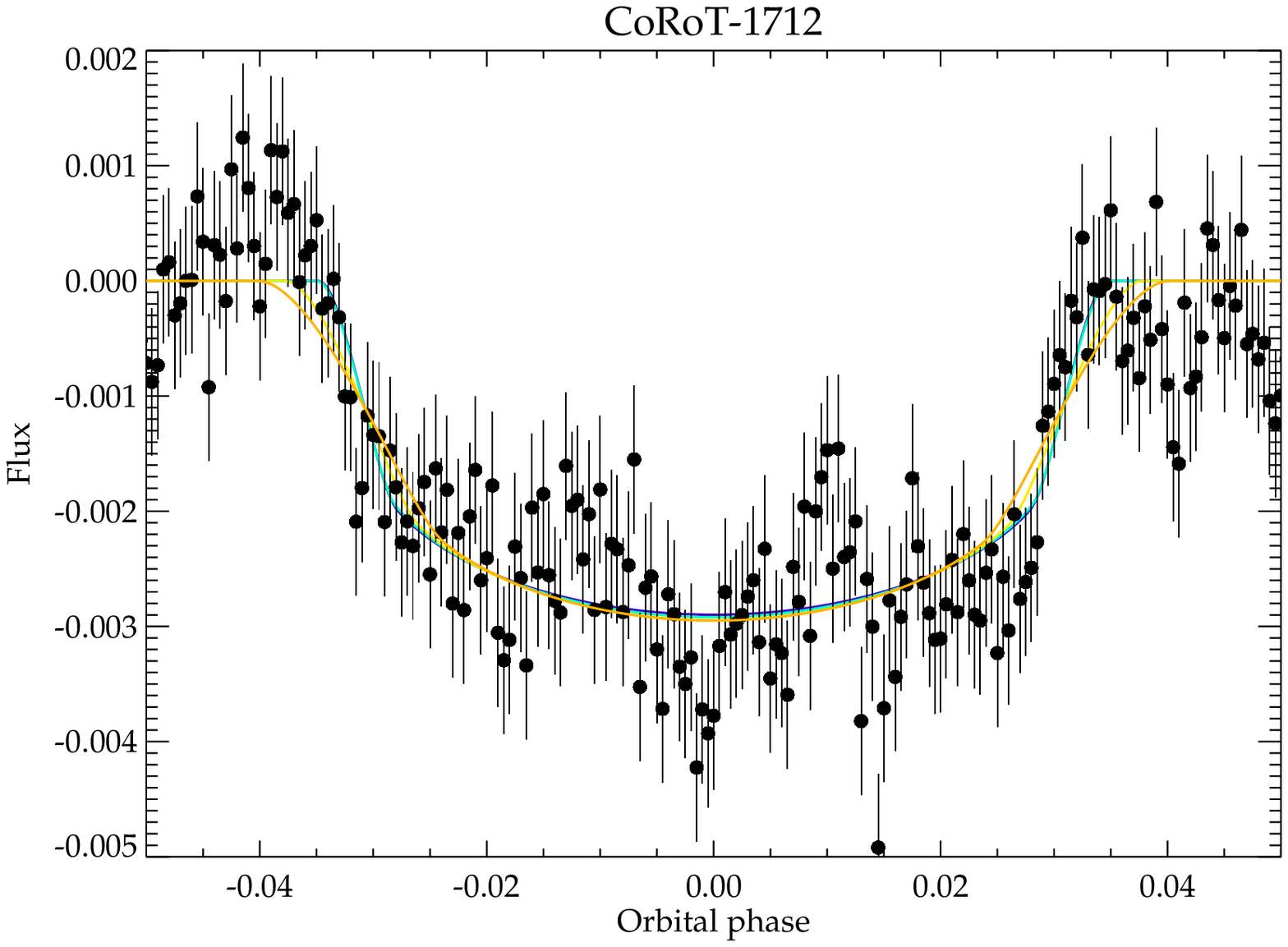}}
{\label{F9}}
\caption{\tiny{For each CoRoT IRa01 candidate: the blended light fraction $k$ versus reduced $\chi^2$ (left panels) and the best fitting blended light models for k=0.2, 0.5, 0.9, and 0.95.}}
\end{figure}

\begin{figure}[h!]
\centering
\subfloat[][CoRoT-4108]{
\includegraphics[width=0.45\textwidth]{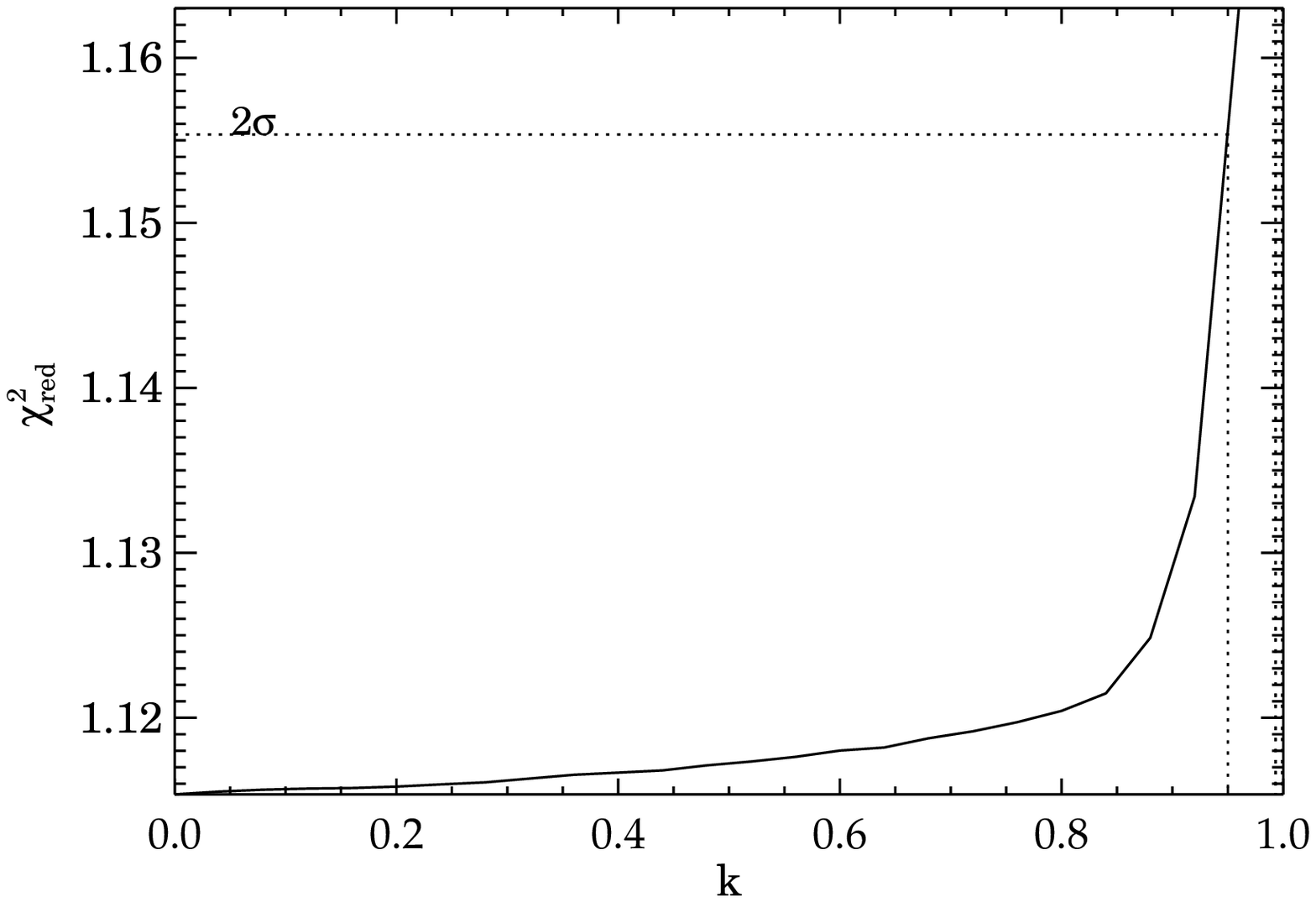}
\includegraphics[width=0.45\textwidth]{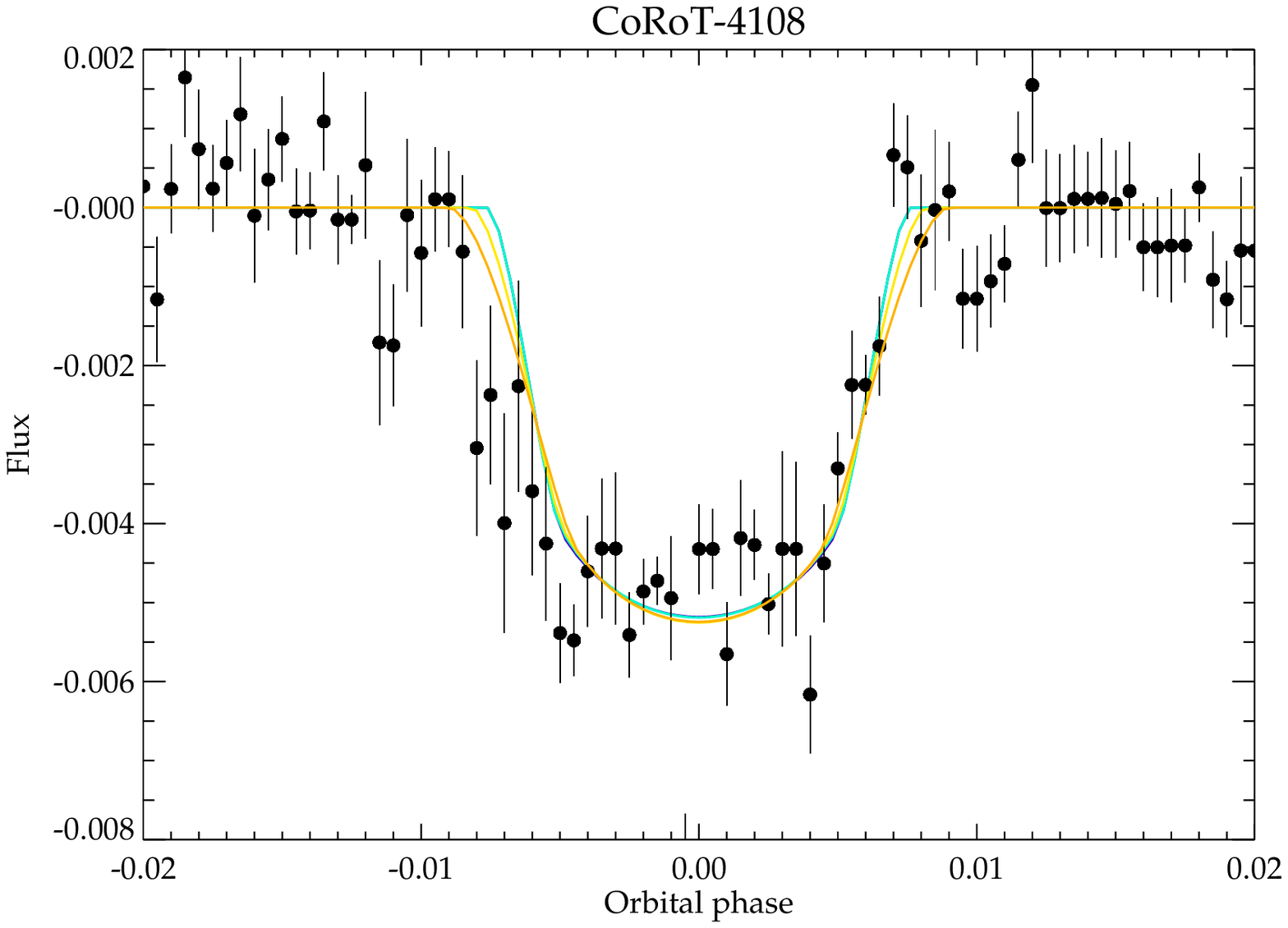}}
\qquad
\subfloat[][CoRoT-4617]{
\includegraphics[width=0.45\textwidth]{EqualParMap16.ps}
\includegraphics[width=0.45\textwidth]{EqualCorotBlend16.ps}}
\qquad
\subfloat[][CoRoT-2430]{
\includegraphics[width=0.45\textwidth]{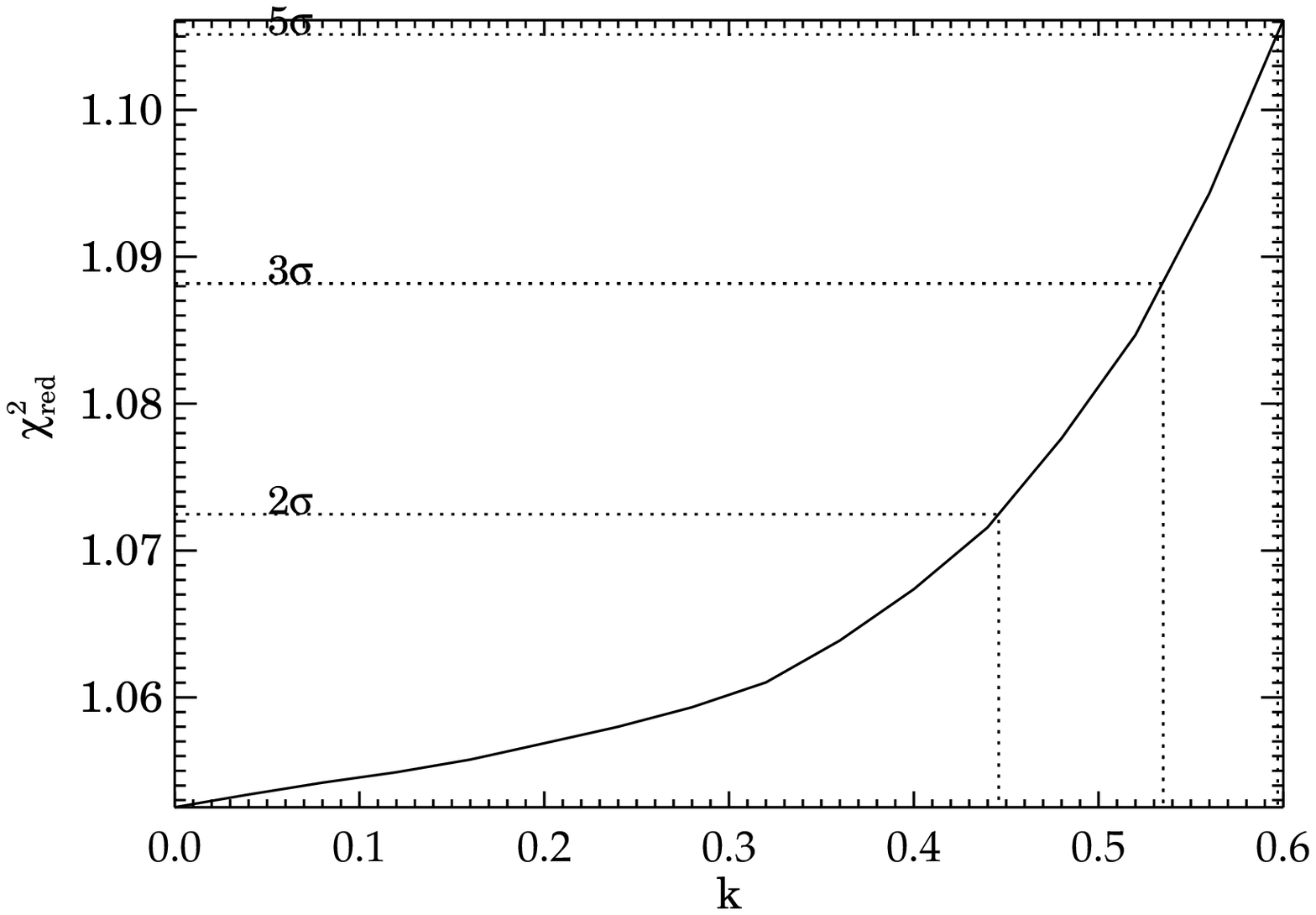}
\includegraphics[width=0.45\textwidth]{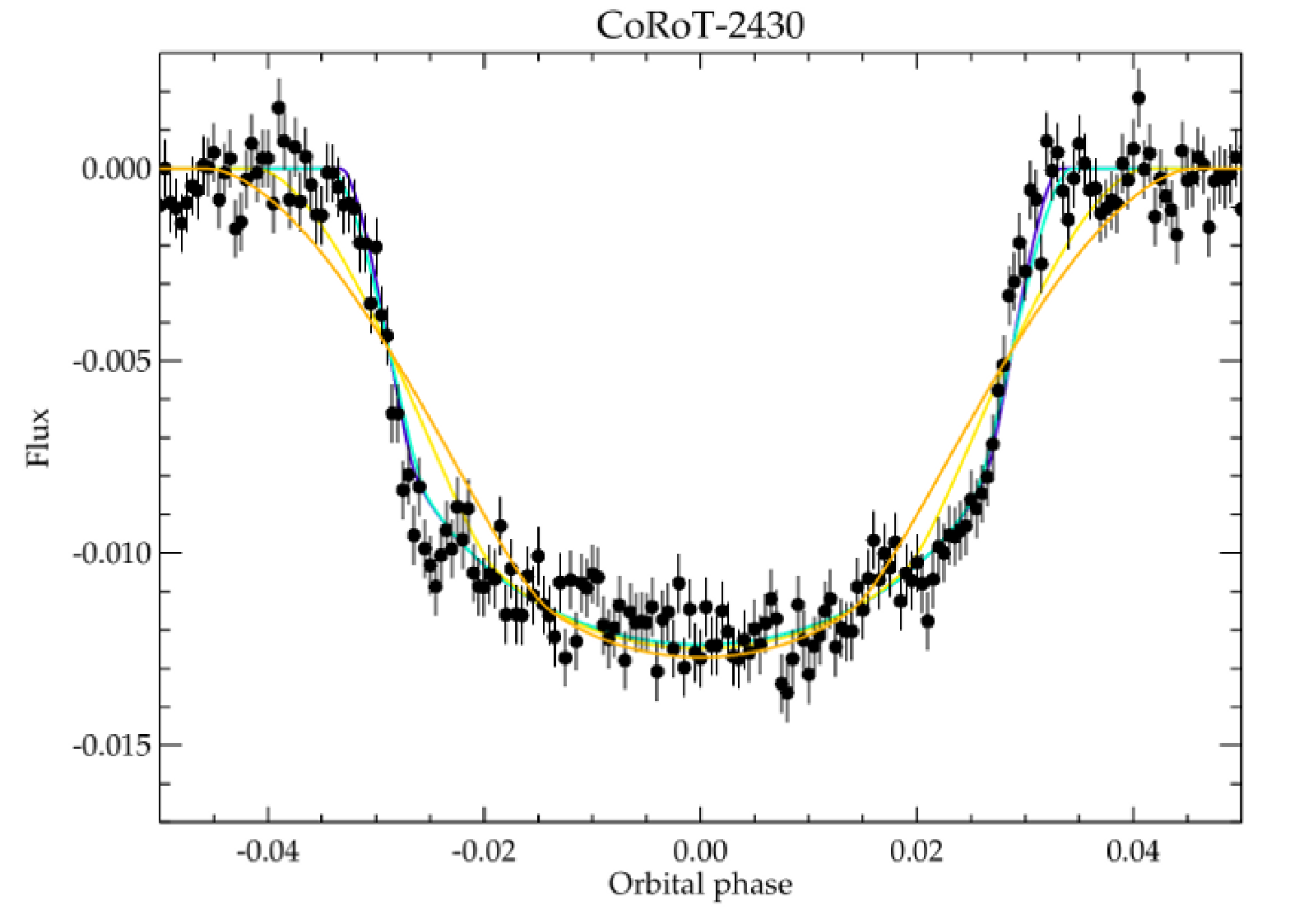}}
\qquad
\subfloat[][CoRoT-4073]{
\includegraphics[width=0.45\textwidth]{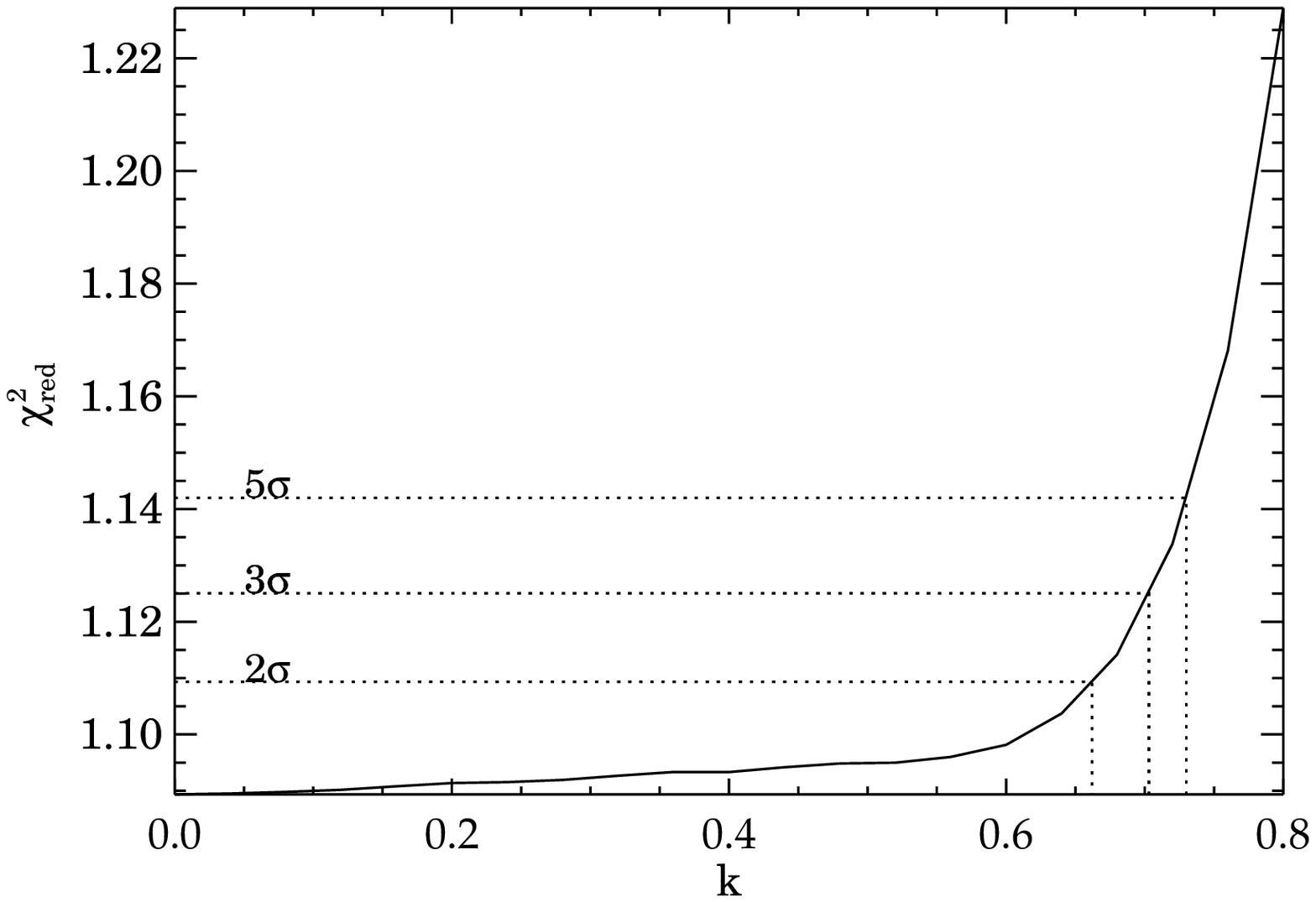}
\includegraphics[width=0.45\textwidth]{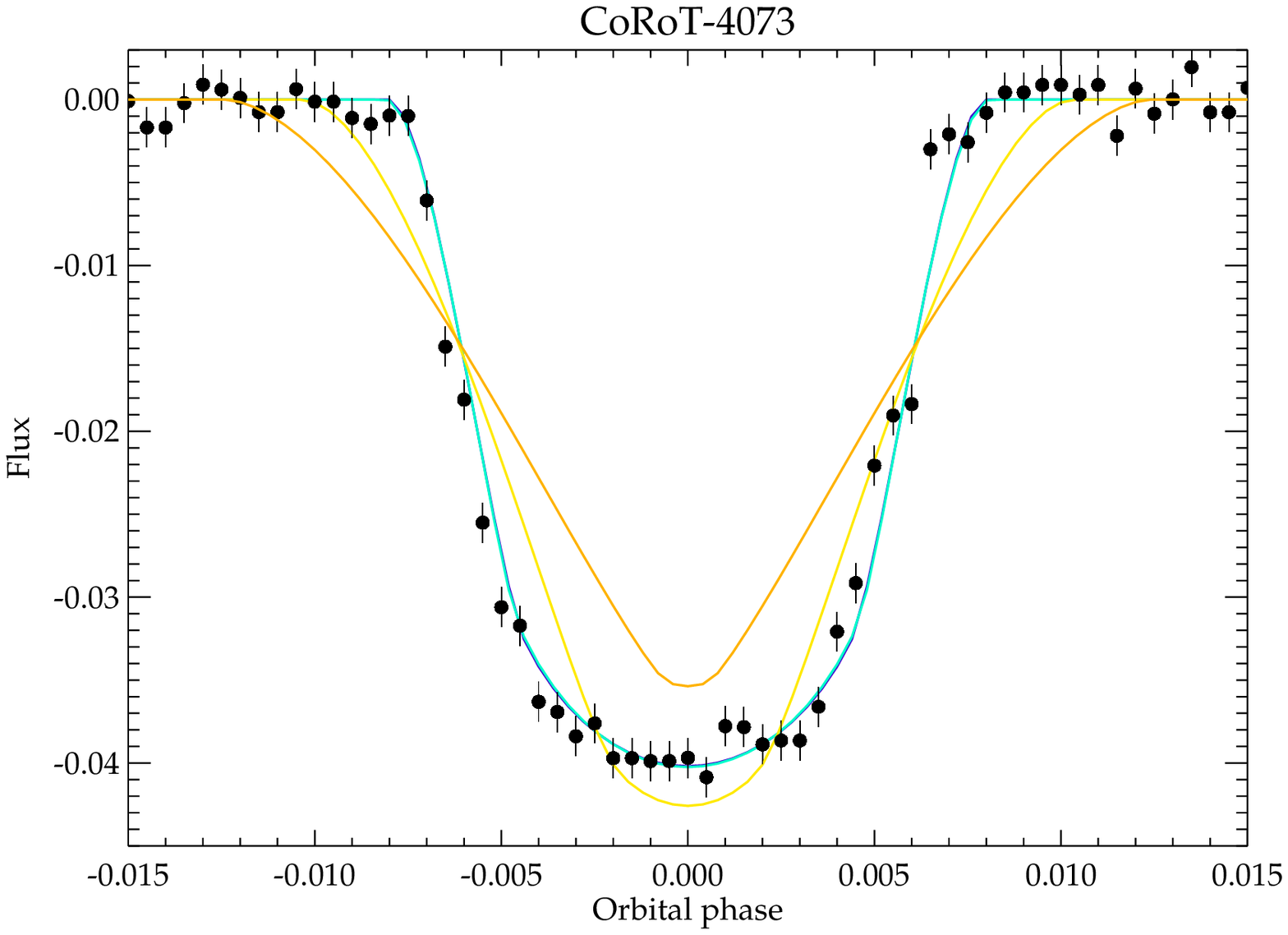}}
\label{Figure10}
\caption{Figure continued}
\end{figure}

\begin{figure}[h!]
\centering
\subfloat[][CoRoT-1736]{
\includegraphics[width=0.45\textwidth]{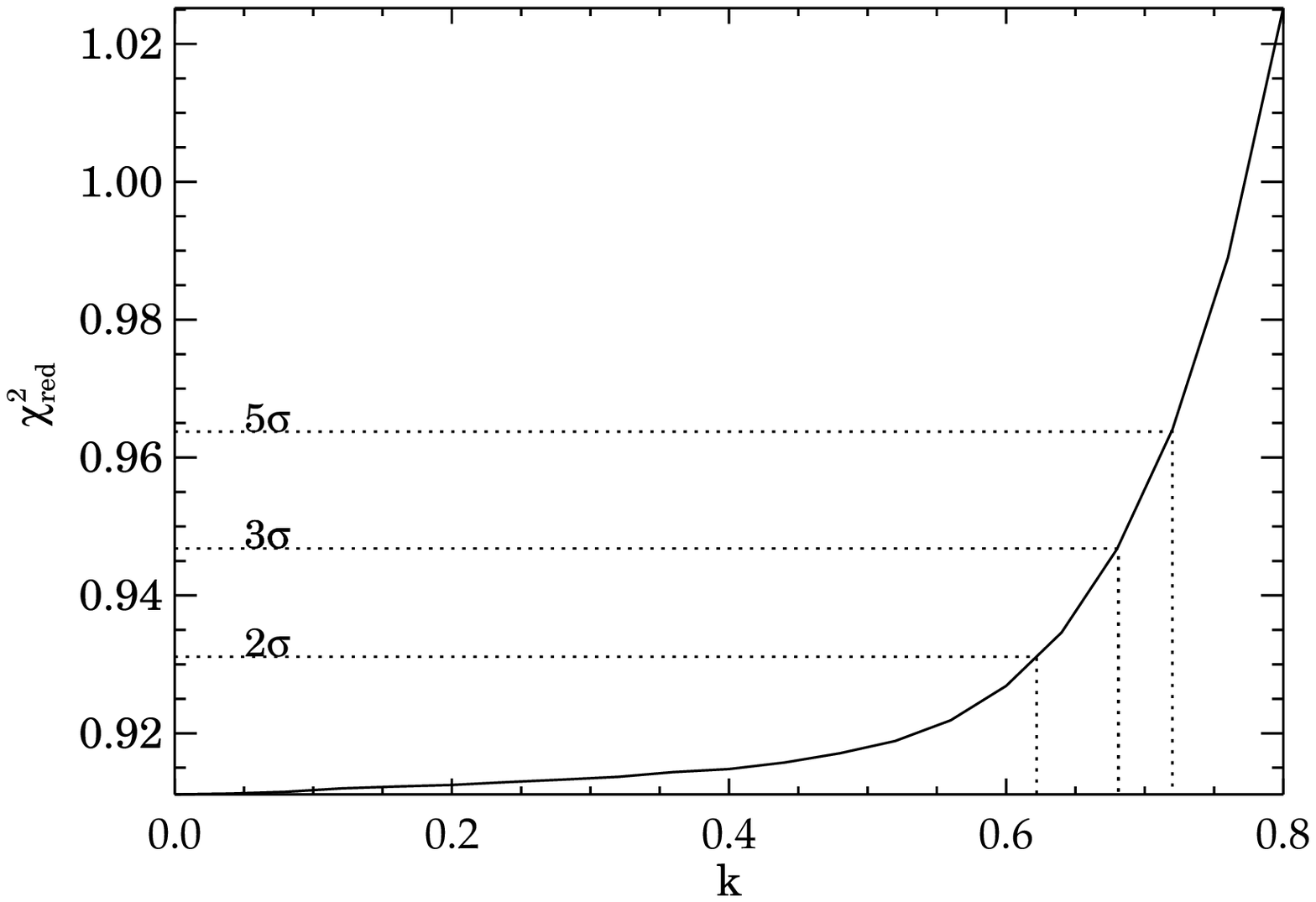}
\includegraphics[width=0.45\textwidth]{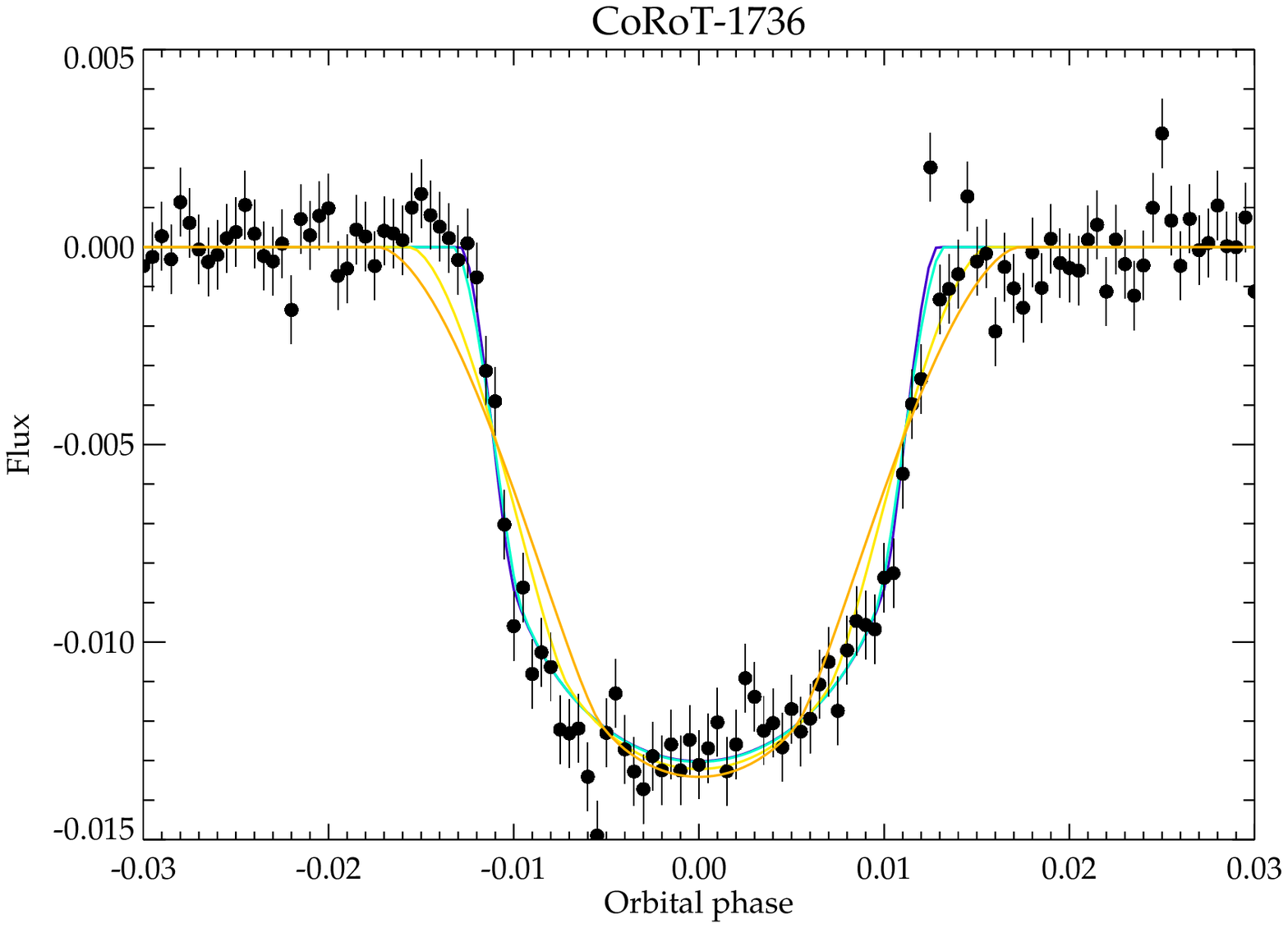}}
\qquad
\subfloat[][CoRoT-3724]{
\includegraphics[width=0.45\textwidth]{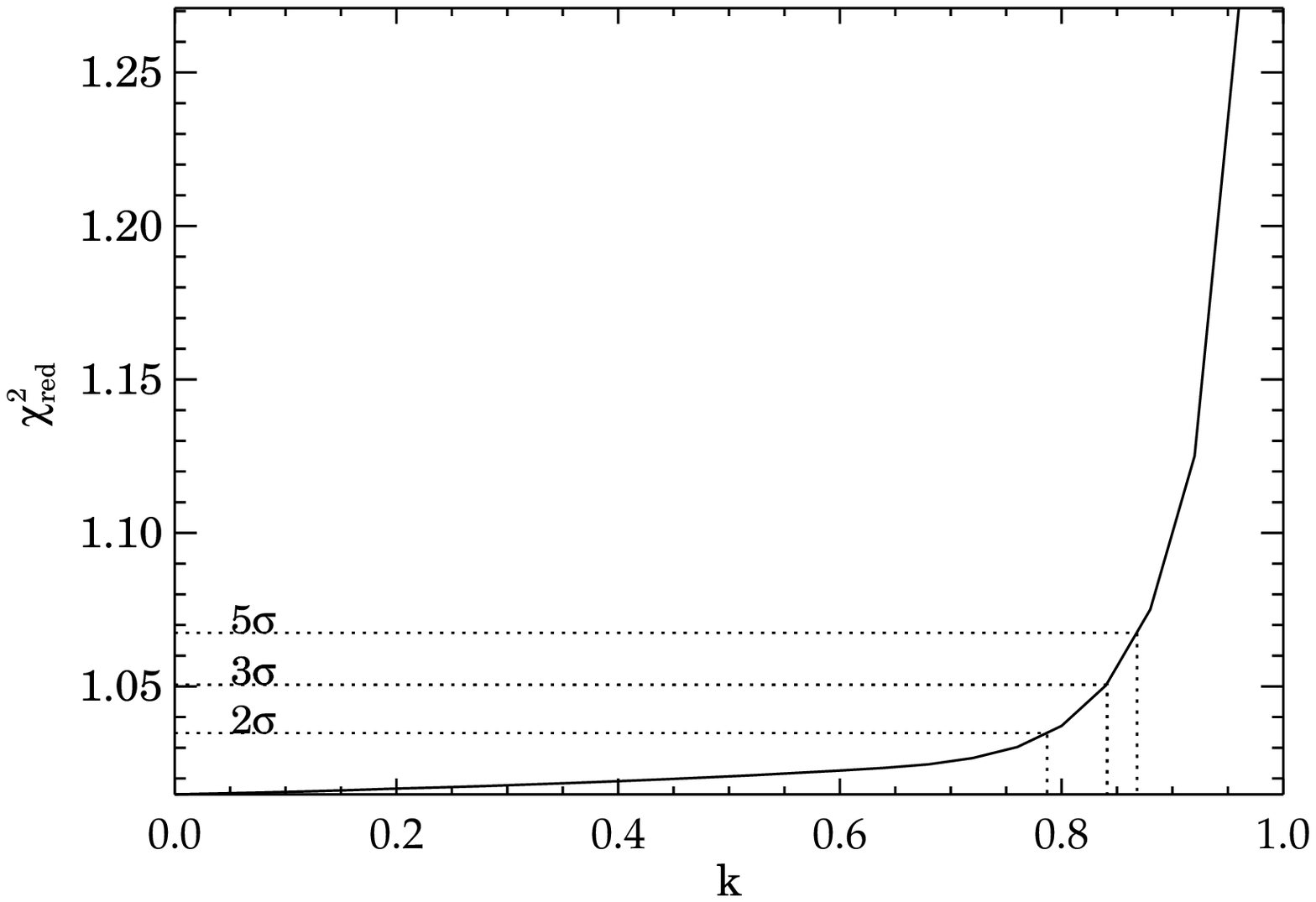}
\includegraphics[width=0.45\textwidth]{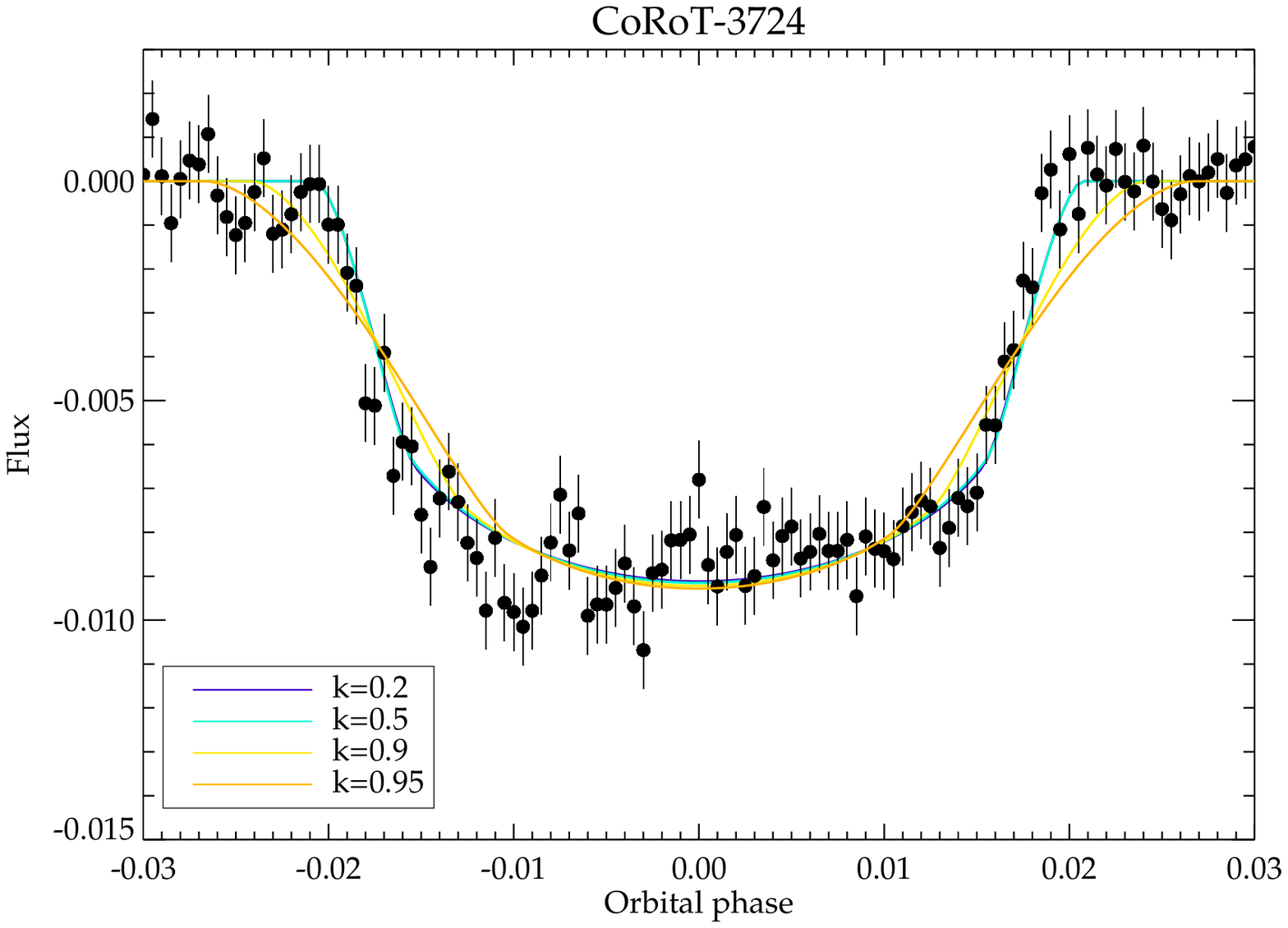}}
\caption{Figure continued}
\label{Figure11}
\end{figure}
\onecolumn
\pagebreak
\begin{table*}[!h]
  {
  \begin{tabular*}{1.0\textwidth}
      {@{\extracolsep{\fill}}cccccc}
       \hline
   WinID+CoRoTID &P(days) &$\left(\frac{r_p}{r_*}\right)$	&$b$	&$\textrm{Log}\left(\frac{\rho_*}{\rho_{\odot}}\right)$(error in $\left(\frac{\rho_*}{\rho_{\odot}}\right) )$	&$\left(\frac{a}{R_*}\right)$\\ 
       \hline \hline
   1319  0102729260&   1.70& 0.17& 1.09(0.011)&-1.36(0.004)& 2.14\\
   1158  0102763847&  10.53& 0.27& 1.10(0.017)& 0.13(0.044)&22.50\\
   0288  0102787048&   7.89& 0.06& 0.90(0.016)&-0.97(0.024)& 8.01\\
   3787  0102787204&   0.86& 0.26& 1.22(0.040)&-1.55(0.002)& 1.17\\
   1857  0102798247&   0.82& 0.07& 0.91(0.013)&-0.78(0.015)& 2.04\\
   4591  0102806520&   4.30& 0.29& 1.21(0.085)&-0.83(0.062)& 5.94\\
   1136  0102809071&   1.22& 0.09& 1.03(0.018)&-1.61(0.002)& 1.41\\
   2430  0102815260&   3.59& 0.10& 0.24(0.107)&-0.81(0.014)& 5.36\\
   0203  0102825481&   5.17& 0.18& 0.62(0.006)& 0.04(0.016)&13.09\\
   1712  0102826302&   2.77& 0.05& 0.60(0.287)&-0.88(0.074)& 4.27\\
   0399  0102829121&  33.06& 0.13& 0.85(0.017)& 0.57(0.243)&67.81\\
   1736  0102855534&  21.72& 0.11& 0.43(0.119)&-1.24(0.009)&12.77\\
   0396  0102856307&   7.82& 0.34& 1.32(0.035)&-1.90(0.030)& 3.90\\
   1126  0102890318&   1.51& 0.14& 0.43(0.017)&-0.16(0.017)& 4.93\\
   0330  0102912369&   9.20& 0.10& 0.18(0.119)&-0.13(0.042)&16.96\\
   2755  0102918586&   4.39& 0.26& 1.01(0.005)&-0.21(0.006)& 9.72\\
   4617  0102753331&  19.76& 0.19& 0.10(0.090)&-1.42(0.001)&10.47\\
   3724  0102759638&  12.33& 0.10& 0.50(0.105)&-1.33(0.008)& 8.17\\
   4290  0102777119&   2.21& 0.14& 1.05(0.010)&-2.77(0.010)& 0.86\\
   4108  0102779966&   7.37& 0.07& 0.80(0.085)&-0.06(0.492)&15.41\\
   1531  0102780627&   2.38& 0.09& 0.91(0.009)&-0.68(0.020)& 4.49\\
   2009  0102788073&  10.85& 0.25& 1.17(0.432)&-1.44(0.045)& 6.88\\
   2774  0102798429&   1.61& 0.29& 1.19(0.133)&-1.32(0.003)& 2.12\\
   3010  0102800106&  23.21& 0.22& 1.00(0.127)&-0.17(0.091)&30.33\\
   4300  0102802430&   5.81& 0.12& 1.00(0.025)&-1.02(0.006)& 6.27\\
   2604  0102805893&   3.82& 0.38& 1.33(0.052)&-1.60(0.009)& 3.04\\
   2648  0102812861&   3.68& 0.10& 0.92(0.070)&-0.82(0.010)& 5.42\\
   2328  0102819021&   4.51& 0.12& 0.97(0.037)&-1.66(0.008)& 3.24\\
   4998  0102821773&  10.08& 0.14& 0.88(0.011)&-0.19(0.067)&17.19\\
   3425  0102835817&   1.19& 0.32& 1.25(0.024)&-1.65(0.008)& 1.34\\
   3854  0102841669&   1.14& 0.05& 0.94(0.050)&-1.40(0.003)& 1.59\\
   3952  0102842120&  13.48& 0.08& 0.85(0.356)& 1.47(0.068)&74.27\\
   1407  0102842459&   5.17& 0.27& 1.02(0.013)& 0.49(0.040)&18.45\\
   2721  0102850921&   0.61& 0.29& 1.18(0.017)&-0.97(0.004)& 1.46\\
   0704  0102855472&   2.16& 0.08& 0.62(0.043)&-1.38(0.005)& 2.45\\
   4073  0102863810&  15.00& 0.18& 0.36(0.036)&-0.08(0.047)&24.40\\
   2329  0102869286&   1.87& 0.13& 1.04(0.432)&-1.41(0.586)& 2.19\\
   3336  0102876631&   1.39& 0.04& 0.84(0.110)&-0.69(0.121)& 3.12\\
   4911  0102881832&   2.17& 0.26& 1.12(0.010)&-1.97(0.013)& 1.57\\
   4339  0102903238&   1.36& 0.07& 1.00(0.126)&-1.63(0.062)& 1.50\\
   4124  0102926194&   1.51& 0.41& 1.37(0.041)&-1.74(0.005)& 1.47\\
   3819  0102932089&   1.57& 0.30& 1.07(0.035)&-0.97(0.012)& 2.73\\
   4467  0102940315&  16.45& 0.19& 0.98(0.049)&-0.86(0.010)&14.25\\
   3856  0102954464&  16.56& 0.49& 1.31(0.047)& 0.55(0.347)&42.02\\  
\hline
\end{tabular*}
    }
  \label{par1}
  \caption{The fitting parameters for our blend models when applied to the CoRoT IRa01 sample, assuming k=0.} 
  \end{table*}

\end{document}